\documentclass[preprint,12pt]{elsarticle}
\usepackage{amssymb}
\usepackage{amsthm}
\usepackage[english]{babel}
\usepackage{pifont}
\usepackage{subcaption}
\usepackage{algorithm}
\usepackage{algpseudocode}
\usepackage{hyperref} 
\usepackage[shortcuts,acronym,nonumberlist,nogroupskip]{glossaries}
\usepackage{nomencl}
\usepackage{tabularray}
\usepackage{adjustbox}
\usepackage[table,xcdraw]{xcolor} 
\newtheorem*{remark}{Remark}

\def\boxit#1{%
  \smash{\color{red}\fboxsep=0pt\llap{\rlap{\fbox{\strut\makebox[#1]{}}}~}}\ignorespaces
}

\makeglossaries 
\newacronym{am}{AM}{active material loss}
\newacronym{adp}{ADP}{approximate dynamic programming}
\newacronym{bm}{BM}{bucket model}
\newacronym{bms}{BMS}{battery management system}
\newacronym{bess}{BESS}{battery energy storage system}
\newacronym{da}{DA}{day-ahead}
\newacronym{der}{DER}{distributed energy resources}
\newacronym{dla}{DLA}{Direct Lookahead}
\newacronym{dso}{DSO}{distribution system operator}
\newacronym{ct}{CT}{continuous-time intra-day}
\newacronym{ems}{EMS}{energy management system}
\newacronym{ema}{EMA}{energy management algorithm}
\newacronym{empc}{eMPC}{economic MPC}
\newacronym{ess}{ESS}{energy storage systems}
\newacronym{ecm}{ECM}{equivalent circuit model}
\newacronym{ev}{EV}{electric vehicle}
\newacronym{fom}{FOM}{full order model}
\newacronym{hess}{HESS}{hybrid energy storage system}
\newacronym{hp}{HP}{heat pump}
\newacronym{hv}{HV}{high voltage}
\newacronym{hvac}{HVAC}{heat, ventilation and air conditioning}
\newacronym{iso}{ISO}{independent system operator}
\newacronym{ida}{IDA}{intra-day auctions}
\newacronym{nmc}{NMC}{nickel manganese cobalt oxides}
\newacronym{mpc}{MPC}{Model Predictive Control}
\newacronym{mces}{MCES}{multicarrier energy systems}
\newacronym{ocv}{OCV}{open-circuit voltage} 
\newacronym{ocp}{OCP}{optimal control problem} 
\newacronym{pei}{PEI}{power electronic interface}
\newacronym{pbrom}{PBROM}{physics-based reduced order model}
\newacronym{rl}{RL}{reinforcement learning}
\newacronym{scm}{SCM}{series connected module}
\newacronym{sei}{SEI}{solid electrolyte interface}
\newacronym{sdp}{SDP}{sequential decision problem}
\newacronym{sddp}{SDDP}{stochastic dual dynamic programming}
\newacronym{soc}{SOC}{State of Charge}
\newacronym{soh}{SOH}{State of Health}
\newacronym{spv}{SPV}{solar photovoltaics}
\newacronym{st}{ST}{solar thermal}
\newacronym{tess}{TESS}{thermal energy storage system}
\newacronym{tso}{TSO}{transmission system operator}
\newacronym{lib}{LIB}{Li-ion batteries}
\newacronym{lto}{LTO}{Lithium-titanate}
\newacronym{lfp}{LFP}{Lithium iron phosphate}
\newacronym{vfa}{VFA}{value function approximation}
\newacronym{umf}{UMF}{Universal Modeling Framework}
\allowdisplaybreaks
\makenomenclature
\renewcommand\nomgroup[1]{%
  \item[\bfseries
    \ifstrequal{#1}{V}{Physics constants}{%
    \ifstrequal{#1}{N}{Number sets}{%
    \ifstrequal{#1}{A}{Optimization, Control \& Dynamics}{%
    \ifstrequal{#1}{C}{Battery Performance Model}{%
    \ifstrequal{#1}{D}{Battery Degradation Model}{%
    \ifstrequal{#1}{E}{Electric Vehicle (EV) Model}{
    }}}}}}%
]}

\nomenclature[A]{$w=[w_{\text{grid}}, w_{\text{loss}}, w_{\text{T}}]$}{Objective function weights.}
\nomenclature[A]{$\tilde{\ \ }$}{Approximated/estimated variable/function.}
\nomenclature[A]{$\overline{\ \ }, \underline{\ \ }$}{Maximum and minimum variable bounds.}
\nomenclature[A]{$^{+}, ^{-}$}{Positive and negative variable components.}
\nomenclature[A]{$\pi$}{policy}
\nomenclature[A]{$t, t', \Delta t$}{Time index, time index within a policy $\pi$ and timestep.}
\nomenclature[A]{$\mathbb{E}_{W}$}{Expectation with respect to the exogenous process $W$}
\nomenclature[A]{$a \in \mathbb{A}$}{Set of energy assets.}
\nomenclature[A]{$X^{\pi}_t$}{Policy function at time $t$}
\nomenclature[A]{$p_{a,t} \in \mathcal{P}$}{Power setpoint for system $a$ at time $t$}
\nomenclature[A]{$C_{\textrm{grid}}$}{Cost of electricity from the grid}
\nomenclature[A]{$C_{\textrm{loss}}$}{Loss cost or penalty}
\nomenclature[A]{$S_{a,t} \in \mathcal{S}$}{State of system $a$ at time $t$}
\nomenclature[A]{$S^M_{a,t}$}{Transition function for asset $a$ at time $t$}
\nomenclature[A]{$W_{t+1}$}{Exogenous information process at time $t+1$}
\nomenclature[A]{$B_{a,t}$}{Belief state for asset $a$ at time $t$}
\nomenclature[A]{$p_{\textrm{T}}$}{Soft-constraint penalty for thermal comfort}
\nomenclature[A]{$\lambda_t$}{Day-ahead buy and sell prices.}
\nomenclature[A]{$T, H$}{Simulation and optimization horizons.}
\nomenclature[A]{$\mathcal{D}_t^{\pi}=[t;t+H^{\pi}]$}{Prediction horizon window at time $t$ for policy $\pi$.}
\nomenclature[A]{$\mathcal{S}^p_t$}{TESS planning terminal set at time $t$.}
\nomenclature[A]{$V^p_t(\cdot)$}{TESS terminal value function at time $t$.}

\nomenclature[C]{$SoC_{b,t}$}{State of Charge of battery $b$ at time $t$}
\nomenclature[C]{$v_{t,b,t}$}{Terminal voltage of battery $b$}
\nomenclature[C]{$t_{0,b}$}{Battery elapsed lifetime}
\nomenclature[C]{$N_{s,b}$}{Number of cells in series}
\nomenclature[C]{$N_{p,b}$}{Number of parallel branches}
\nomenclature[C]{$\eta_c$}{Coulombic efficiency}
\nomenclature[C]{$Q_{b,t}$}{Battery capacity at time $t$}
\nomenclature[D]{$SoH$}{State of Health of the battery}
\nomenclature[D]{$i_{\text{loss},b,t}$}{Total degradation current}
\nomenclature[D]{$i_{\textrm{SEI},b,t}$}{SEI side reaction current}
\nomenclature[D]{$\eta_{k,b,t}$}{SEI kinetic overpotential}
\nomenclature[C]{$OCV_{b,t}$}{Open-circuit voltage at time $t$}
\nomenclature[D]{$OCV_{n,b,t}$}{Open-circuit voltage of anode}
\nomenclature[D]{$z_{b,t}, z_{b,0\%}, z_{b,100\%}$}{Lithium stoichiometry at time $t$, and limits at 0 and 100\%}
\nomenclature[D]{$i_{\textrm{AM},b,t}$}{Active material loss current}
\nomenclature[D]{$\delta_{\textrm{SEI},b,t}$}{SEI layer thickness}
\nomenclature[D]{$\varepsilon_{e,b,t}$}{Electrolyte volume fraction}

\journal{Advances in Applied Energy}

\begin{document}

\begin{frontmatter}



\title{Building Seasonal Highways for Residential Energy Hubs: Sizing, planning and operating thermal energy storage}


\author[inst1]{Darío Slaifstein}
\author[inst2]{Mohammad Khosravi}
\author[inst1]{Gautham Ram Chandra Mouli}
\author[inst1]{Laura Ramirez-Elizondo}
\author[inst1]{Pavol Bauer}

\affiliation[inst1]{organization={DC Systems, Energy Conversion \& Storage, Electrical Sustainable Energy Department, Delft University of Technology},
            addressline={Mekelweg 8}, 
            city={Delft},
            postcode={2628}, 
            state={Zuid-Holland},
            country={Netherlands}}
 \affiliation[inst2]{organization={Delft Center for Systems and Control, Delft University of Technology},
            addressline={Mekelweg 2}, 
            city={Delft},
            postcode={2628}, 
            state={Zuid-Holland},
            country={Netherlands}}

\begin{abstract}
The operation of residential energy hubs with multiple energy carriers (electricity, heat, mobility) poses a significant challenge due to the energy storage differences in time-constants, round-trip efficiencies and self-discharge rates. Usually, thermal storage exhibits flexibility in yearly planning optimizations or long-term scenarios. However, as optimization horizons shrink (1-48hs) so does their supplied value due to the lower round-trip efficiencies. To avoid this early depletion during operation this paper proposes a data-driven highway to steer the short-term daily control towards long-term optimality. The proposed methodology also presents how to optimally size the thermal storage and avoid yearly simulations and how all of this is related to nonlinearities in the daily operation. The presented framework links seasonal and daily optimizations through dynamic terminal sets and value functions. The seasonally-aware nonlinear economic model predictive controller achieves the most balanced performance, with the second best mean grid cost of all MPCs at -\texteuro 209. It also achieves better battery degradation control than its linear counterparts (between 26-34\%) and the best thermal comfort of the nonlinear benchmarks. Nevertheless, the data-driven seasonal highway restrains the ability to control battery degradation and slightly increases computational time.
\end{abstract}

\begin{graphicalabstract}
\includegraphics[width=\textwidth]{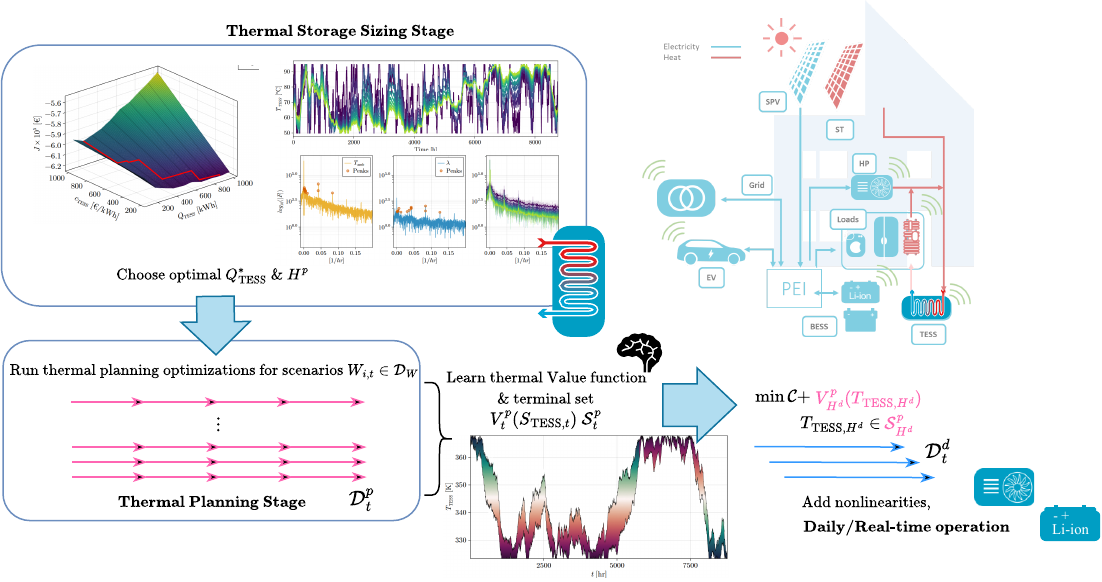}
\end{graphicalabstract}

\begin{highlights}
\item Data-driven dynamic terminal set and value function steer short-sighted nonlinear economic model predictive controller (eMPC) towards seasonal optimal.
\item Seasonally and ageing aware nonlinear eMPC beats other benchmarks in mean grid cost, beats linear alternatives with less battery degradation and maintains building thermal comfort. All at approximately 7.4 seconds per optimization.  
\item Seasonal steering the thermal energy storage limits the ability of the eMPC to control battery degradation and slightly increases computational time.
\end{highlights}


\end{frontmatter}


\section{Introduction}

Nowadays, building electrification constitutes a significant challenge within the energy transition. The electrification of all domestic and commercial applications is driven by the energy efficiency of electrical alternatives such as \ac{hp} and \ac{ev}. Nevertheless, beyond the first phase of replacing gas boilers with \ac{hp}s; \ac{tess}, that stores energy in heat, appears as an interesting solution to achieve lower costs and seasonal flexibility \cite{Alpizar-Castillo2024ModellingHouse, Geidl2007, Glucker2024OptimalVector}. Electrical storage systems are also becoming interesting for the residential sector \cite{Li2019}. Not only due to the increasing presence of \ac{ev}s but due to the rapid rise of \ac{bess} solutions to complement rooftop \ac{spv}. 

To operate such systems, \ac{ems} are designed to sequentially decide power setpoints, considering power limits, costs, and user demands (electricity, heat, transport). This is called \ac{sdp} and aims to choose power for storage devices (\ac{bess}, \ac{tess}) and heat injection by the \ac{hp}, seeking to minimize costs and capacity. Within it, \ac{tess} operation is challenging due to its large size and low round-trip efficiency when compared to \ac{bess}.

In this context, the question of how much storage is needed and what storage supports a given service (power, heating, etc.) is critical to avoid over-investment. For example, does the \ac{bess} completely absorb the uncertainty of the electric load? How much thermal storage is needed for seasonal variations? Is there a synergy between the \ac{ev} and \ac{tess}? What is the role of passive thermal storage (building inertia) and active thermal storage (\ac{tess})?

The integration of heat and power in residential systems has been studied thoroughly in the literature. Focusing on the integration of sizing, planning \& operation, the works of Geidl \cite{Geidl2007} and  Hafiz \cite{Hafiz2019EnergyApproach} present the now standard approaches. The traditional standard is a hierarchical optimization strategy. First, the top layer decides for the size of assets to be installed (number of PV panels, charger size, \ac{bess} capacity, etc.) assuming a certain operation (fixed number of cycles, average irradiance, etc.) and financial variables (interest rate, depreciation, unit cost, etc.). Later, seasonal scenarios are run to determine long-term optimal patterns. Finally, this optimal long term pattern is passed as a reference to the daily operation layer which could be a tracking \ac{mpc} or PID controller. However this architecture can be suboptimal in settings with high short-term volatility, in which \ac{empc} presents increased flexibility \cite{Kohler2024AnalysisOverview}.

In optimal control settings unwanted effects take place when combining \ac{tess} with electrical storage systems, such as \ac{bess} or \ac{ev} with bi-directional charging. Among them, two effects/problems are prominent in the operational layer when carelessly implementing \ac{empc} solutions. First, as the finite optimization horizon $H$ of the \ac{empc} increases, the closer the finite-horizon dispatch gets to the infinite-horizon dispatch for a given storage system \cite{Rawlings2022ModelEdition, Grune2017NonlinearAlgorithms, Kohler2024AnalysisOverview}. The smallest optimization horizon for which the finite-horizon dispatch converges to the infinite-horizon dispatch is called the minimum forecast horizon \cite{Prat2024HowProblems}. This minimum forecast horizon grows with storage size and with smaller round-trip efficiencies \cite{Prat2024HowProblems}. As such, thermal seasonal storage has long (weeks, months, etc) minimum forecast horizons and electric storage has small (hours, days) minimum forecast horizons, because the roundtrip efficiency of \ac{bess} is quite larger than of \ac{tess}. Secondly, when two or more storage systems are combined the optimization horizon $H$ has to be larger than the largest minimum forecast horizon among storage systems. Otherwise, storage systems with minimum forecast horizon much larger than the $H$ will get depleted without being charged again  \cite{Darivianakis2017AManagement, Brosius2026MitigatingFunctions}.


The coordination between different storage devices has been studied extensively in the literature. The most common methods are \ac{adp} and \ac{sddp}, originally designed for hydro-thermal economic dispatch \cite{Pereira1991Multi-stagePlanning, Rosemberg2022AssessingModels}. Hole et al. \cite{Hole2023CapacityProgramming} presents a classic capacity expansion planning problem for the New Zealand power system. However, a major problem with dynamic programming approaches is the curse of dimensionality and their inability to handle non-convex dynamics \cite{Powell2022}. To avoid dynamic programming and reduce the number of decision variables, Feng \cite{Feng2024HybridApproach} transforms the time-domain constraints to the frequency domain where the main natural and excitation modes are identified and used to optimize over. A second option to reduce the number of variables is to use multi-horizon \ac{mpc} to extend the optimization horizon but extending the sampling times as the horizon grows \cite{Behrunani2024DistributedHubs}. Bender's decomposition and/or bi-level optimization is also common to size and coordinate assets in different time-scales \cite{Yu2025CoordinatedNetworks, Chen2025CapacityComfort}. A secondary problem with traditional decomposition strategies,  like SDDP or Bender's decomposition, is that they are generally limited to linear or quadratic optimization problems \cite{Rahmaniani2017TheReview}. For other convex problems, specific conditions have to be met for the master problem and subproblems to converge to optimal solutions. This rules out incorporating nonlinear dynamics directly into the optimization constraints, because the information communicated between master and sub-problem loses its value. Finally, Darivianakis \cite{Darivianakis2017AManagement} and Brosius \cite{Brosius2026MitigatingFunctions} propose to learn seasonal information from year-long optimizations and pass it to the operational layer that handles short-term uncertainty. However, it bears the question of the relationship between the size of the storage and the length of the planning/seasonal optimization horizons. In summary, to exploit the \ac{tess}, one must rigidly track a seasonal reference calculated off-line \cite{Hole2023CapacityProgramming, Rosemberg2022AssessingModels}, or be limited to simplified linear dynamics leaving potential savings untapped \cite{Feng2024HybridApproach, Behrunani2024DistributedHubs, Yu2025CoordinatedNetworks, Chen2025CapacityComfort, Darivianakis2017AManagement, Brosius2026MitigatingFunctions}, and only use seasonal thermal storage sizes \cite{Brosius2026MitigatingFunctions, Darivianakis2017AManagement}.

Following the above discussion, this paper presents a novel applied framework to size and coordinate electro-thermal energy storage systems with different time-flexibilities in order to effectively control residential energy hubs with minimum operational cost. To address the issues mentioned above, the contributions of the proposed framework are listed below.
\begin{itemize}
    \item we develop a generalized methodology to determine the size of thermal storage $Q_{\text{TESS}}$ and planning horizon $H^p$ needed beyond seasonal storage, including weekly and monthly sizes.
    \item  we propose an approach that includes objectives with soft-constraints.
    \item we introduce the expansion of the operational layer by including nonlinear continuous dynamics, excluding mixed-integer constraints.
\end{itemize}


\section{Thermal-sizing \& planning framework}
\label{sec:modeling}


\begin{figure}[tb]
    \centering
    \includegraphics[width=\linewidth]{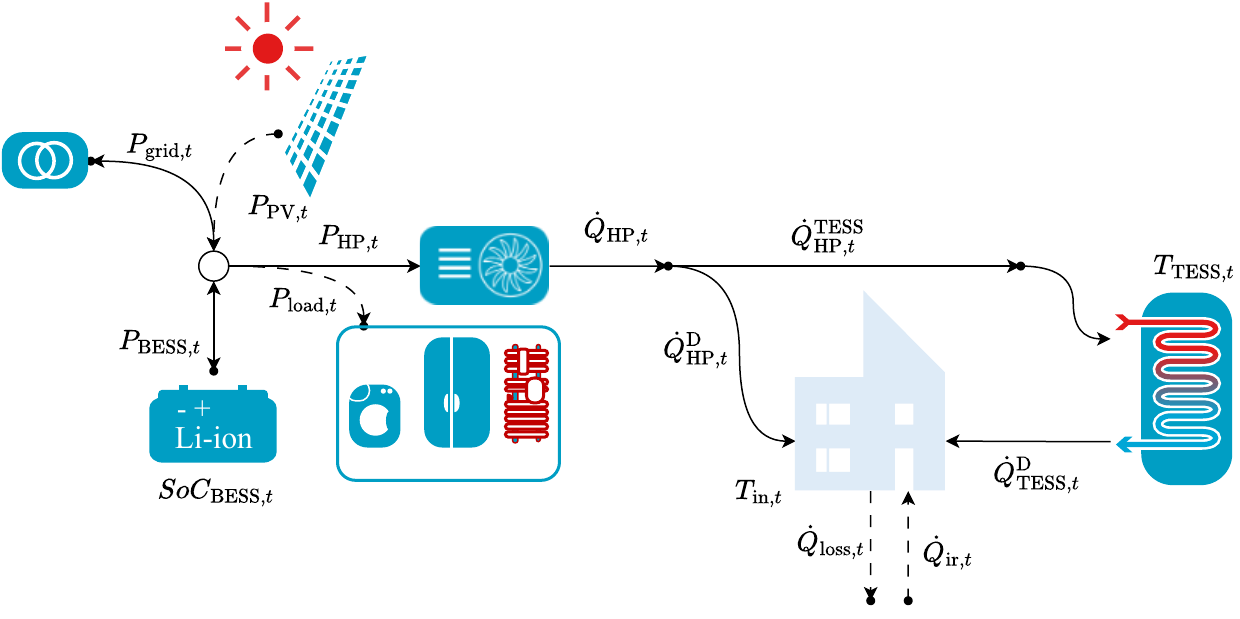}
    \caption{System flows and balances. Full lines indicate controllable flows $p_{a,t}$ and dashed lines show exogenous/uncontrollable disturbances $W_{t}$.}
    \label{fig:sysModel}
\end{figure}
\begin{figure*}[tb!]
    \centering
    \includegraphics[width=\textwidth]{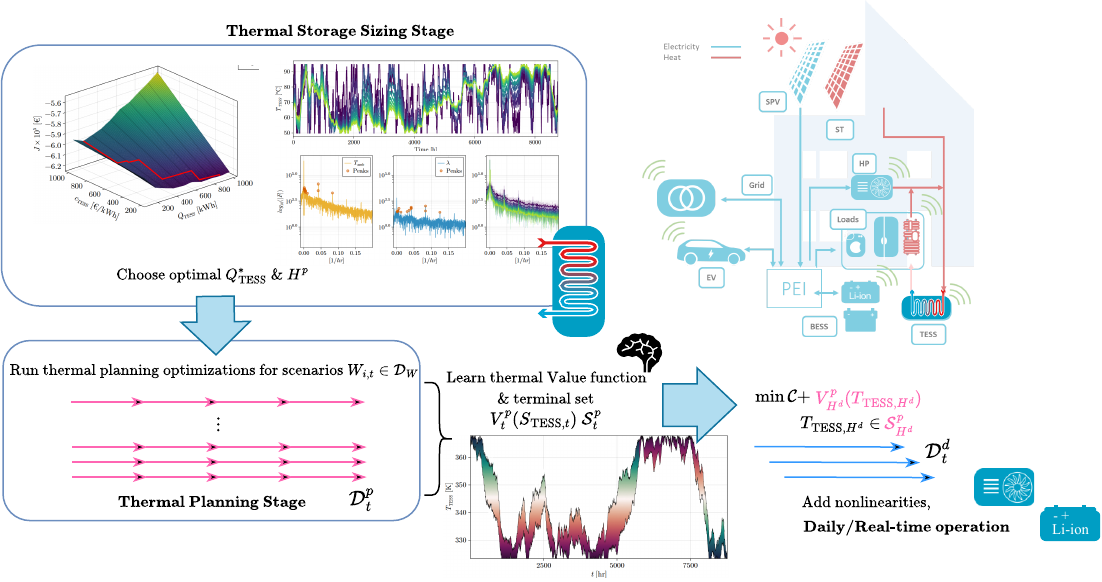}
    \caption{Planning and sizing workflow. The first stage determines the size of the \ac{tess} $Q_{\textrm{TESS}}$ and the minimum planning horizon $H^p$. These are used to run stochastic planning optimizations to learn the terminal set $\mathcal{S}_t^p$ and value function $V_t^p(\cdot)$. The two components are added to the final daily operation policy.}
    \label{fig:decomp}
\end{figure*}

The primary goal is to minimize the mean monetary stage cost, i.e. $\mathbb{E}[\mathcal{C}]$, over the exogenous random process $W \in \mathcal{D}_W$, by choosing an optimal size $x_{a}^*$ and power dispatch $p_{a,t}$. The stage cost function $\mathcal{C}$ comprises the net grid cost $C_{\text{grid}}$, the battery capacity fade cost $C_{\text{loss}}$, and a thermal comfort penalty $p_{T}$. The grid cost aims to represent the economic operation of the building, the battery degradation is calculated by the cell ageing current $i_{\text{loss}}$ multiplied by the number of cells in a pack $N_{\text{s}}N_{\text{p}}$ and finally the thermal comfort is a soft-constraint of the indoor temperature $T_{\text{in},t}$ outside of the bounds $[\underline{T}_{\text{in}},\overline{T}_{\text{in}}]$ when the building is occupied $\mathcal{O}_t$ \cite{Slaifstein2026SequentialMPC}. The components of the objective are:
\begin{subequations}
\begin{align}
    & V_f=\sum_{a} c_a x_a,\ \\
    & \mathcal{C} = C_{\text{grid}}+C_{\text{loss}}+p_{T},\ \\
    & C_{\text{grid},i}=w_{\text{grid}}\sum_{t=0}^{T}{\lambda_{t} P_{\text{grid},t} \Delta t},\ \\
        & C_{\text{loss}}=w_{\text{loss}} c_{\text{loss}} \sum_{t=0}^{T}{N_{\textrm{s}} N_{\textrm{p}}i_{\text{loss},t} \Delta t},\ \\
    & p_{T} = w_{T} \sum_{t=0}^{T}s_{T_{\text{in}},t}\mathcal{O}_{t} \Delta t,\
\end{align}
\end{subequations}

\noindent{The general stochastic sizing problem is:}
\begin{subequations}\label{eq:fullProb}
    \begin{align}
        \min_{x_a^*, p_{a,t}^* \in \mathcal{X}} \quad &  V_f+\mathbb{E}_{W} \left[ \mathcal{C} \right]\\
        \text{s.t.}\ \ \quad & S_{a,t+1} = S_{a}^M\left(S_{a,t}, x_a^*, W_{t+1} \right)  \quad \forall a \in \mathbb{A}, \ W_{t} \in \mathcal{D}_W \\
        & x_a^*, p_{a,t}^* = X^{\pi}_t(S_{a,t}) \in \mathcal{X} \quad \forall a \in \mathbb{A} , \ W_{t} \in \mathcal{D}_W \\
        & S_{a,i,t} \in \mathcal{S} \quad \forall a \in \mathbb{A}, \ W_{t} \in \mathcal{D}_W\\
        \text{with}\ \quad &\ \mathbb{A}=\{\text{SPV}, \text{grid}, \text{BESS}, \text{HP}, \text{TESS}\}\,.
    \end{align}
\end{subequations}

\noindent{where $S_{a,i,t}$ is the state vector, $x^*_a$ is the optimal sizes vector, $p_{a,i,t}^*$ are the optimal power flows, $W_{i,t+1}$ is an exogenous process sampled trajectory that introduces new information after making a decision. Objective weights $w$ scale and blend the different objectives into $\mathcal{C}$. The mappings $S_{a,t}^M(\cdot)$, and $X^{\pi}_t(\cdot)$ are the transition function and optimal policy, respectively. The first is a set of equations describing the states and parameter evolution, and the second is the algorithm that finds the optimal decision. The subindex $a \in \mathbb{A}$ corresponds to the assets shown in Fig. \ref{fig:sysModel} and $i \in [1,N_s]$ is the sample index. The control actions/decisions are in the set $\mathcal{X}$, the states in $\mathcal{S}$, time is in the domain $t\in \mathcal{D}_t=[0,T]$, and exogenous disturbances come from the random field $\mathcal{D}_W$.}

The general sizing and scheduling problem in Eq. \ref{eq:fullProb} becomes untractable when: long optimization horizons are used, battery ageing is integrated in $S^M_{a,t}(\cdot)$ \cite{Slaifstein2026SequentialMPC}. The reader might be tempted to solve the problem in a sequential decision manner, applying deterministic economic \ac{mpc} \cite{Slaifstein2026SequentialMPC}. However, if the optimization horizon $H^d$ of the individual optimizations is too short, the low efficiency long-term storage will be depleted and unused.

A similar problem is introduced in \cite{Darivianakis2017AManagement}, where a deterministic \ac{mpc} is enhanced with a seasonal value function $V^s_t$ and terminal set $\mathcal{S}^s_t$ learned from past scenarios. This idea is extended in this paper by including the sizing of the long term storage, the design methodology of the long-term stochastic optimization, the inclusion of soft-constraints in the objective and the extension to nonlinear economic \ac{mpc} due to the incorporation  of physics-based battery ageing and heat-pump coefficient of performance. This is summarized in the optimization framework presented in Fig. \ref{fig:decomp}.

First, the \ac{tess} capacity $Q_{\text{TESS}}$ is chosen based on its fixed costs (interest rate and capital expenditure) and their operational costs (yearly optimizations). Later, for the chosen $Q^*_{\textrm{TESS}}$ size, its temperature response $T_{\textrm{TESS},t}$ to the exogenous processes of ambient temperature $T_{\text{amb},t}$ and energy prices $\lambda_t$ is used to choose the minimum planning horizon $H^p$ for long term optimizations. Secondly, the capacity $Q^*_{\text{TESS}}$ and planning horizon $H^p$ are used to run stochastic scenarios and determine a dynamic terminal set $\mathcal{S}^p_t$ and terminal value functions $V^p_t(\cdot)$. Finally, these two new components are added to the standard daily operational \ac{empc}. Conceptually, these two new components act like a highway for a car, where the terminal set $\mathcal{S}_t^p$ is analogous to road and lanes and the value function $V_t^p(\cdot)$ indicates which lanes attract more value. Within this highway the \ac{empc} decides in real-time its position (power flows). Each block of the framework will be further explained in the remaining sections.
\begin{figure*}[bth]
    \centering
    \includegraphics[width=\linewidth]{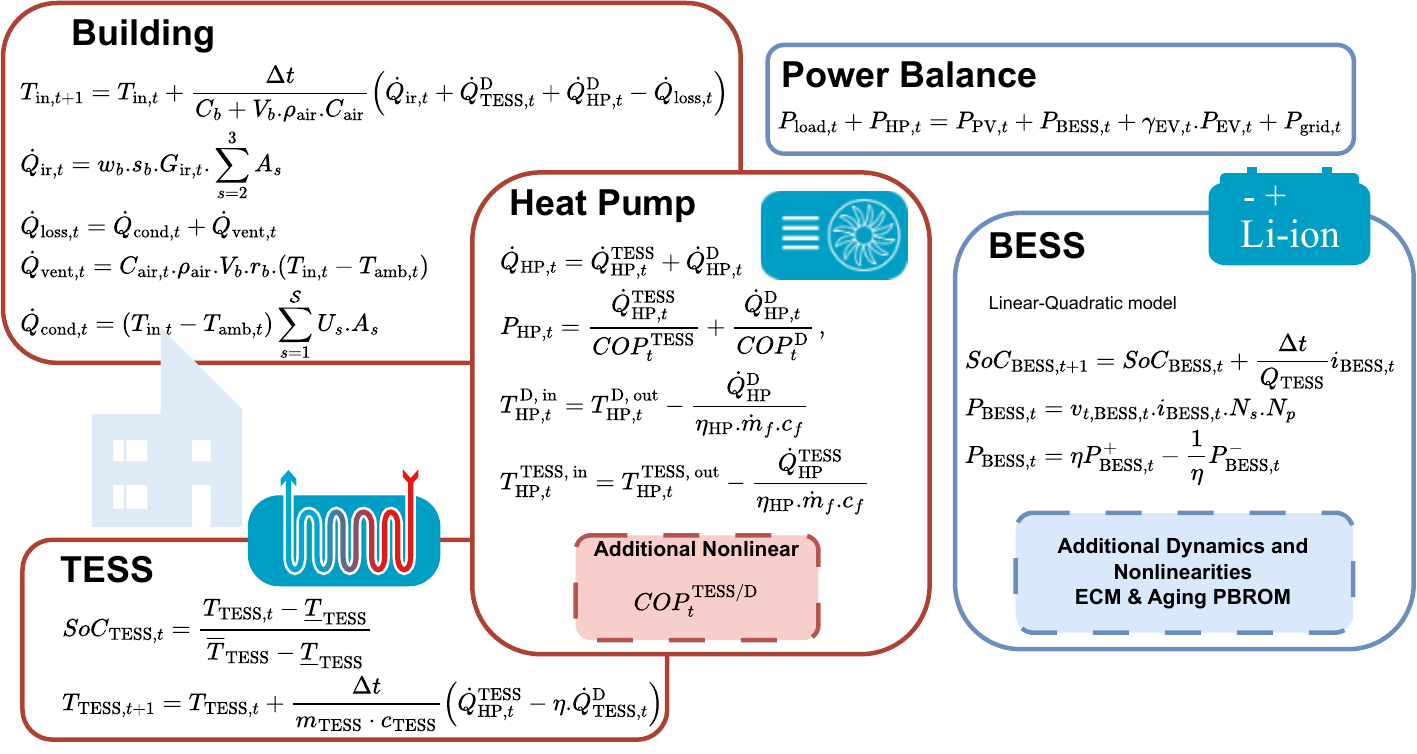}
    \caption{Constraint sets and models in each stage. The basic models for sizing and planning appear with a white background. For short-term operation additional nonlinearities are included (battery degradation and heat-pump conversion).}
    \label{fig:model_stages}
\end{figure*}

\subsection{Dynamic behavior \& Operational constraints}

For the presented energy hub, the variables of the optimal control problems are the power flows $p_{a,t}$ and states $S_{a,t}$ of asset $a$ at time $t$.  The upper and lower limits  $\overline{P},\ \underline{P},\ \overline{S}$ and $\underline{S}$  bound the powers and states. The exogenous information is $W_{i,t+1} = \left[\lambda, G_{\text{ir}}, P_{\text{load}}, T_{\text{amb}}, \mathcal{O} \right]_{i,t+1}^T$. The state variables are $S_{a,i,t}= \left[  SoC_{\text{BESS}}, T_{\text{TESS}}, T_{\text{in}}\right]_{i,t}^T$\,. The decision variables are the power and heat flows $ p_{i,t}=\left[ P_{\text{grid}}, P_{\text{BESS}}, P_{\text{HP}}^{\textrm{e}},\dot{Q}_{\text{HP}}^{\text{D}}, \dot{Q}_{\text{HP}}^{\text{TESS}}, \dot{Q}_{\text{TESS}}^{\text{D}}  \right]_{i,t}^T$. Finally, the only asset being sized is the \ac{tess} capacity $x_a \rightarrow Q_{\textrm{TESS}}$. 

The device models used in each stage of the framework change along the way. From the least detailed linear dynamic models in the sizing and planning stages, to the detailed non-linear physics-based battery and heat-pump models in the operational layer. A visual summary of the models is presented in Fig. \ref{fig:model_stages}. The first two stages of sizing and planning use simplified linear models whereas the daily operation uses the nonlinear dynamic models presented in \cite{Slaifstein2026SequentialMPC}.

The energy hub operational constraints are represented by the set $\text{EH}_t$ and defined by the exogenous processes $W_{t+1}$ and internal parameters. The set only includes equalities and inequality constraints, without dynamic equations correlating time steps. The two carriers' flow balances link the demands, the generators, and the storage devices. The electrical and thermal systems are linked through the \ac{hp}. The balances and operation constraints define state set $\mathcal{S}$ and decisions $\mathcal{X}$ and are combined in the set:
\begin{equation}
    \begin{aligned}
     & \text{EH}_{t}\left(W_{t+1}\right)=\begin{cases}
            \textbf{PB}_{p} p_{a,t}+\textbf{PB}_{W} W_{t+1}= 0, \\
            \textbf{O}_{S} S_{a,t}+\textbf{O}_{p} p_{a,t}+\textbf{O}_{W} W_{t+1} \leq h, \\
        \end{cases}
    \end{aligned}
\end{equation}
\noindent{where matrices $\textbf{PB}_{p}, \textbf{PB}_{W}, \textbf{O}_{S}, \textbf{O}_{p},$ $\textbf{O}_{W}$ and $h$ are all known and summarized in \ref{app:models}.}

\subsection{Uncertainty characterization}

\begin{figure}[bt]
    \centering
    \includegraphics[width=\linewidth]{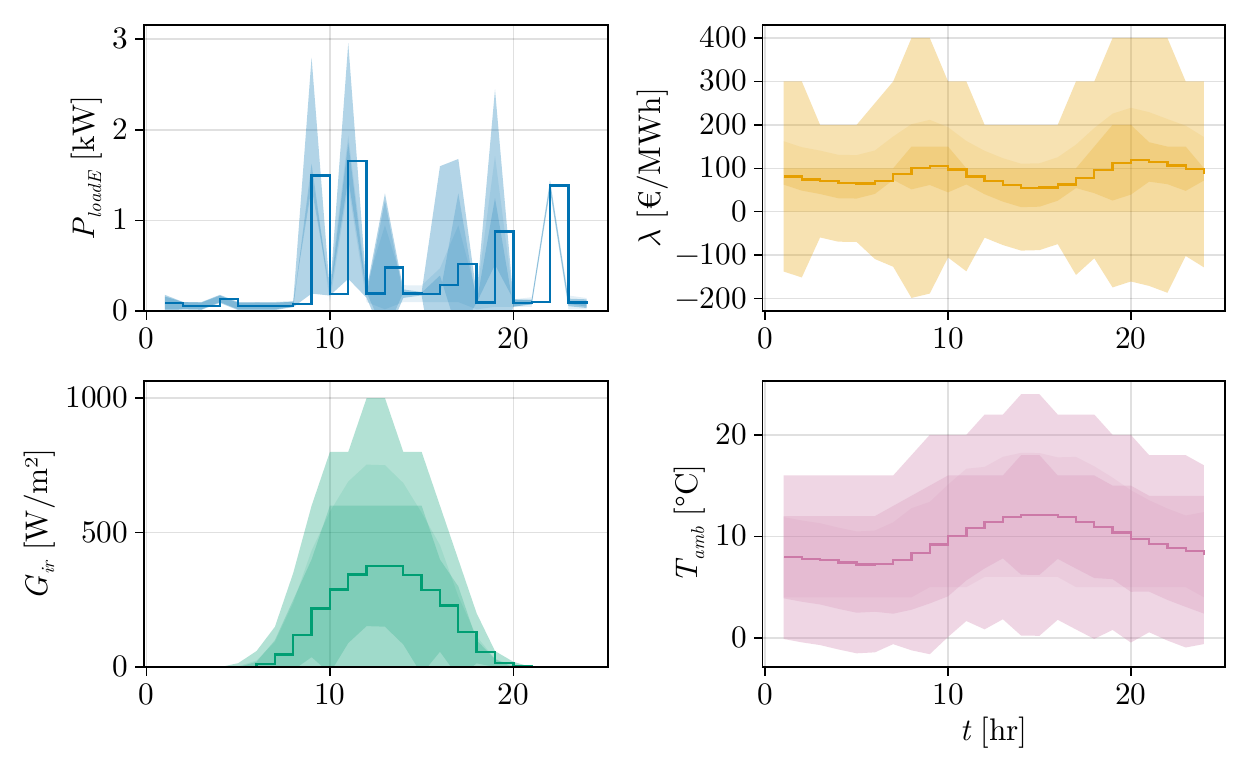}
    \caption{Daily patterns of exogenous information $W_{t+1}$. The bands represent confidence intervals at 50\%, 75\% and 95\%, with the line showing the daily mean.}
    \label{fig:dailyBox}
\end{figure}

To construct the scenarios/realizations of $\mathcal{D}_W$ historical data and probability distributions were taken from public sources \cite{KNMI, 2023EPEXServices, elaad, Pipattanasomporn2014LoadOpportunities} and The Green Village \cite{GreenVillage}. Figure \ref{fig:dailyBox} presents the daily medians and confidence bounds (50\%, 75\% and 95\%) built from the historical data. For the energy prices $\lambda_t$, the day-ahead auction prices for the Netherlands from 2020 to 2024 are taken \cite{2023EPEXServices}. The ambient temperature $T_{\text{amb},t}$ and solar irradiance $G_{\text{ir},t}$ are taken from KNMI for the period 2005 to 2023 \cite{KNMI}. The electric load of the house is built through the method proposed in \cite{Pipattanasomporn2014LoadOpportunities}. The availability $\gamma$ and occupancy $\mathcal{O}$ profiles are built as specified in \cite{Slaifstein2024StochasticHubs, Damianakis2023Risk-averseConsumption}.

To generate stochastic scenarios the historical data is used to build year-long hourly histograms, i.e. 8760 histograms for each of the mentioned exogenous parameters in $\mathcal{D}_W$. For the prices $\lambda_t$, temperature $T_{\textrm{amb},t}$, and irradiance $G_{\textrm{ir},t}$ this is critical because they contain seasonal patterns. Since it does not include heating or cooling the electric load $P_{\textrm{load},t}$, the \ac{ev} availability $\gamma_t$ and building occupancy $\mathcal{O}_t$  don't present seasonal patterns, thus the daily histograms in 15-min resolution are enough to generate as many scenarios as needed.

\subsection{Long-term storage sizing}\label{ssec:sizing}

Now, with the statistical representation of the random field $\mathcal{D}_W$ ready we pass to its exploitation. To size the long-term storage \ac{tess}, several yearly deterministic optimizations are run for various $Q_{\text{TESS}}$ values. The system is optimized using only linear dynamic models. The sizing policy $X^y_t$ consists on solving the deterministic yearly optimization for different $Q_{\textrm{TESS}}$:
\begin{subequations}\label{eq:yearOpt}
    \begin{align}
         \min_{ p_{a,i,t}^* \in \mathcal{X}} \quad & \mathcal{C}^y\\
   \text{s.t.}\ \quad & S_{a,t+1} = \textbf{A} S_{a,t}+\textbf{B} p_{a,t}^*+\textbf{C} W_{t+1}  \\
        & \left(S_{a,t}, p_{a,t}\right) \in \text{EH}_t\left(W_{t+1}\right) \\
        & \quad \forall a \in \mathbb{A},\ \forall t \in \mathcal{D}_{t}^y.
    \end{align}
\end{subequations}

\noindent{The transition function $S^M_t$ describes device dynamics and is linearized in the system $\textbf{A},\ \textbf{B},\ \textbf{C}$.} 

The total annual cost $J^y$, is defined by the fixed investment cost $V_f^y$ and the annual operational cost $\mathcal{C}^y$. In the first place, the investment is defined by the unit cost of storage $c_{\textrm{TESS}}$ in [\texteuro/kWh], the project lifetime $LFT$ and interest rate $ir$.  They are all related as
\begin{subequations}\label{eq:yCosts}
    \begin{align}
        & J^y = V_f^y + \mathcal{C}^y, \\
        & V_f^y = c_{\textrm{TESS}}\frac{ir(1+ir)^{LFT}}{(1+ir)^{LFT}-1}Q_{\textrm{TESS}},\\
        & \mathcal{C}^y = \sum_{t=0}^{H^y} \Big( \lambda_{t} P_{\text{grid},t}
          + w_T s_{T_{in},t}\mathcal{O}_{t}+ \notag \\
        & \qquad\qquad w_p\left\lVert P_{\text{HP},t}\right\rVert_2^2
          + w_p\left\lVert \dot{Q}_{\text{TESS},t}^{D}\right\rVert_2^2 \Big)\Delta t.
    \end{align}
\end{subequations}

The yearly optimization of Eq. \ref{eq:yearOpt} is solved for the mean scenario $\mathbb{E}[W_{t+1}]$ and for several $Q_{\text{TESS}}$ values. The reduced objective is to minimize the grid cost $C_{\text{grid}}$, and ensure thermal comfort $p_{T}$. The last two components minimize the thermal actions and ensure sparsity in the solution. The $\mathcal{C}^y$ is monotonically decreasing along $Q_{\text{TESS}}$, larger sizes less $C_{\textrm{grid}}$. However the investment $V^y_f$ linearly increases with size, causing a trade-off point between $V_f^y$ and $\mathcal{C}^y$. The optimal size $Q^*_{\textrm{TESS}}$ is a function of the interest rate $ir$ and the unit cost $c_{\textrm{TESS}}$. Thus, for a given $ir$ and $c_{\textrm{TESS}}$ the optimal capacity $Q^*_{\textrm{TESS}}$ is determined.

On the other hand, to estimate the minimum planning horizon necessary $H^p$ the relationship between the \ac{tess} (natural time-constant) and the exogenous information (disturbances) is analyzed. The exogenous $W_{t}$, with freq. spectrum $W_k$, disturbs the \ac{tess} which reacts with its freq. response $T_{\textrm{TESS},k}$. With this response $T_{\textrm{TESS},k}$, and the linearized dynamics of Eq. \ref{eq:yearOpt} the minimum planning horizon is estimated as $H^p \approx 3\tau^p$.
\begin{remark}
    As a sanity check, a practical way of verifying $H^p$ is: run a rolling horizon optimization with the same scenario it was defined. The trajectories should closely match those from the yearly optimization of $H^y$, showing performance convergence.
\end{remark}

\subsection{Thermal Planning layer}

Having defined the long-term storage size $Q_{\text{TESS}}^*$ and planning horizon $H^p$, a planning optimal control problem is built and solved under $N_s$ scenarios $W_{i, t+1} \in \mathcal{D}_W$. The stochastic planning policy $X^p_{i,t}$ solves the \ac{ocp}: 
\begin{subequations}\label{eq:planocp}
    \begin{align}
    \min_{ p_{a,i,t}^* \in \mathcal{X}} \quad &  \mathcal{C}_i^p\\
        \text{s.t.}\ \quad & S_{a,i,t+1} = \textbf{A} S_{a,i,t}+\textbf{B} p_{a,i,t}^*+\textbf{C} W_{i,t+1}  \\
        & \left(S_{a,i,t}, p^*_{a,i,t}\right) \in \text{EH}_t\left(W_{i,t+1}\right) \\
        & \quad \forall a \in \mathbb{A}, \ i \in \mathcal{N}_s,\ \forall t \in \mathcal{D}_{t}^p.
    \end{align}
\end{subequations}

The planning \ac{ocp}, Eq.\ref{eq:planocp}, differs from the yearly optimization, Eq. \ref{eq:yearOpt}, in the fact that the optimization horizon $H^p < H^y$ resulting in a smaller computational load. This is the main goal of choosing $Q^*_{\textrm{TESS}}$ and $H^p$, avoiding time-consuming yearly optimizations and reducing the risk of untractability. The planning model is used in two ways, to learn the dynamic terminal set and to learn the value function.

To learn the dynamic terminal set $\mathcal{S}^p_t$ the planning model, Eq. \ref{eq:planocp}, the model is solved for scenarios $i\in \mathcal{N}_s$ as many rolling horizons $H^p$ as necessary. The terminal set is then the 95\% confidence interval of the resulting $T_{\textrm{TESS},i,t}$. 

To learn the planning value function $V^p_t(\cdot)$ the planning \ac{ocp}, Eq. \ref{eq:planocp}, is solved for multiple initial conditions $T_{\textrm{TESS},t_0}$ and multiple $\mathcal{N}_s$ realizations for each planning horizon $H^p$. The $V_t^p(\cdot)$ is defined using the previously obtained $c_{\text{grid},i,t}^p$  as:
\begin{subequations}
    \begin{align}
            V_t^p(T_{\textrm{TESS},t})=\min_{ p_{a,i,t}^* \in \mathcal{X}} \quad &  \mathbb{E}_W \left[\sum_{t=0}^T c_{\text{grid},i,t}^p \right]\\
        \text{s.t.}\ \quad & S_{a,i,t+1} = \textbf{A} S_{a,i,t}+\textbf{B} p_{a,i,t}^*+\textbf{C} W_{i,t+1}  \\
        & \left(S_{a,i,t}, p^*_{a,i,t}\right) \in \text{EH}_t\left(W_{i,t+1}\right) \\
        & \quad \forall a \in \mathbb{A}, \ i \in \mathcal{N}_s,\ \forall t \in \mathcal{D}_{t}^p.
    \end{align}
\end{subequations}

\subsection{Operational layer}

The operational layer is an economic \ac{mpc} where all carriers and storage dynamics are described with their detailed non-linear model $S^M_t$. The operational \ac{empc} now includes the \ac{bess} physics-based battery ageing and nonlinear coefficient-of-performance for the \ac{hp} \cite{Slaifstein2026Ageing-awareSystems, Slaifstein2026SequentialMPC}.  The operational policy $X^d_t$ has an optimization horizon $H^d=24$hs, and the dynamic \ac{ocp} is defined as:
\begin{subequations}\label{eq:sagempc}
    \begin{align}
    \min_{p_{a,i,t}^{*} \in \mathcal{X}} \quad & \mathcal{C}_i+V^p_{H^d}\left(T_{\textrm{TESS},tt'_0+H^d}\right)\\
    \text{s.t.} \quad & S_{a,i,tt'+1} = S_{a,t}^M\left(S_{a,i,tt'}, p_{a,i,tt'}^*, W_{i,tt'+1} \right) \\
        & \left(S_{a,i,tt'}, p_{a,i,tt'}\right) \in \text{EH}_t\left(W_{i,t+1}\right) \\
        & T_{\text{TESS},tt'_0+H^d} \in \mathcal{S}^p_{tt'_0+H^d}\\
        & S_{\text{BESS},tt'_0} = S_{\text{BESS},tt'_0+H^d}\\
        & \forall a \in \mathbb{A}, \ i \in \mathcal{N}_s,\ \forall t \in \mathcal{D}_{t},\ \forall t' \in \mathcal{D}_{t}^d
    \end{align}
\end{subequations}


\section{Case Study \& Results}
\label{sec:results}
The system is composed of a 5kWp \ac{spv}, a 12.5kW/15.6kWh \ac{bess} with NMC cells, a 4kWe heat pump, a 2.5kWp electrical load, and a 17kW LV grid connection.  The models were developed using Julia \cite{Julia}, JuMP \cite{JuMP} and InfiniteOpt \cite{InfiniteOpt}, with KNITRO as a solver \cite{KNITRO}. The offline stochastic planning simulations were run on a Intel i7 CPU 4 cores 16GB of RAM. Due to restrictions with KNITRO, the \textit{LP} \ac{mpc}s were run using 8 CPUs of 4GB of RAM each and the \textit{NL} \ac{mpc}s were run using 1 CPU of 32GB of RAM. The latter simulations were done using the high-performance-computing (HPC) cluster DelftBlue \cite{DHPC2024DelftBlue2}. The source-code for reproducing the results can be found in the following repository \hyperlink{https://github.com/DarioSlaifsteinSk/SAGeMPC}{SAGeMPC}.  

Although the general mathematical formulation incorporates \ac{ev} charging dynamics and bidirectional Vehicle-to-Grid (V2G) flexibility into \ac{ems} \cite{Slaifstein2026SequentialMPC}, the main numerical benchmark focuses on the core electro-thermal interaction between the BESS, TESS, and Heat Pump. Isolating these assets prevents stochastic user mobility noise (unpredictable \ac{ev} arrival/departure times \cite{Slaifstein2024StochasticHubs}) from interfering the performance evaluation of seasonal thermal steering and electrochemical aging reduction. Nevertheless, the reduced-order electro-thermal aging models derived herein remain fully compatible with joint V2G-TESS operation.

\subsection{Sizing Thermal Storage}
\label{sec:rSizing}

\begin{figure*}[tbh]
    \centering
    \begin{subfigure}{0.45\linewidth}
        \includegraphics[width=0.95\linewidth]{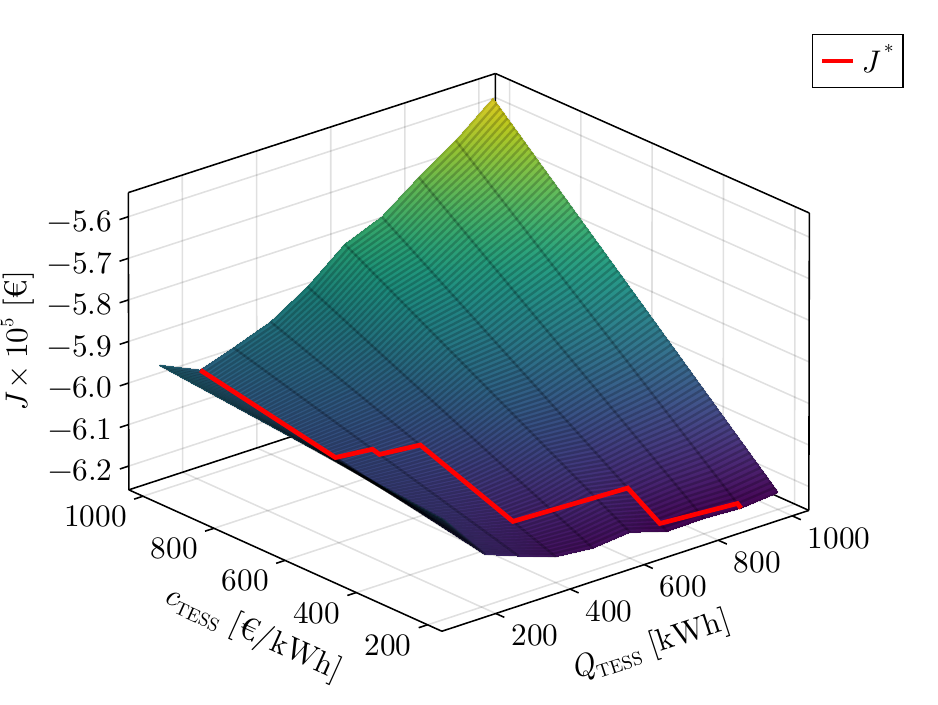} 
        \caption{$J^y(Q_{\text{TESS}},c_{\text{TESS}}, 5\%)$.}
        \label{fig:costInvQtess}
        \end{subfigure}
        \begin{subfigure}{0.45\linewidth}
        \includegraphics[width=0.95\linewidth]{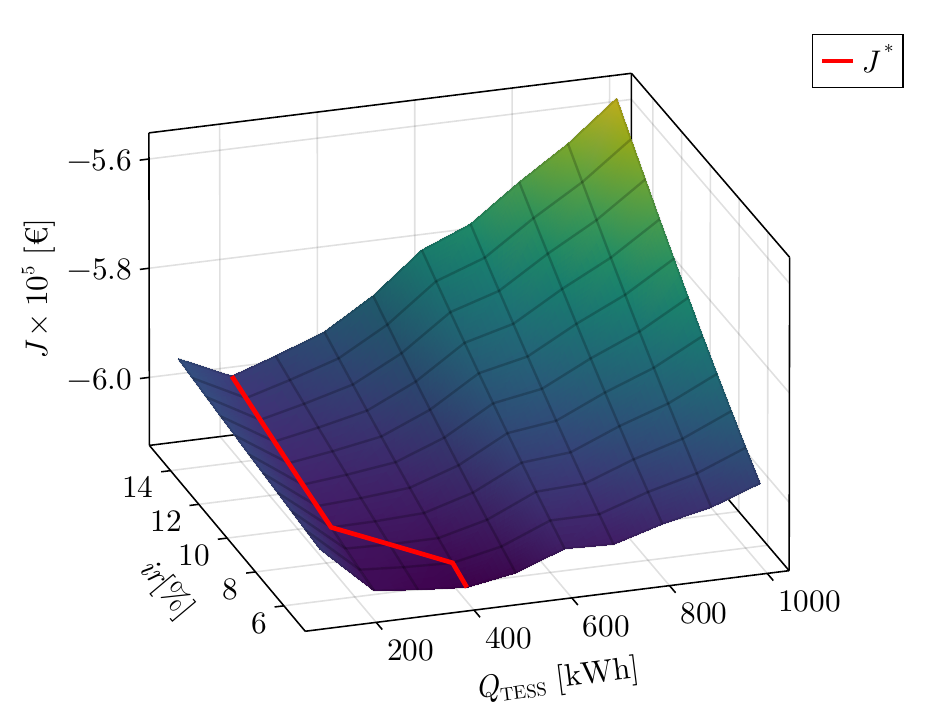}
        \caption{$J^y(Q_{\text{TESS}}, 500\text{\texteuro/kWh},ir)$.}
        \label{fig:costirQtess}
    \end{subfigure}
    \caption{Yearly cost function for various \ac{tess} sizes, $J^y(Q_{\text{TESS}},c_{\textrm{TESS}},ir)$.}
    \label{fig:costQtess}
\end{figure*}
Following Section \ref{ssec:sizing}, the yearly optimization of Eq. \ref{eq:yearOpt} is solved for the mean scenario $\mathbb{E}[W_{t+1}]$ and for $Q_{\text{TESS}}\in [100, 2100]$kWh and reduced objective $J^y$. The total cost $J^y$ as a function of $Q_{\textrm{TESS}}$ and financial parameters ($c_{\textrm{TESS}},\ ir$) is presented in Fig. \ref{fig:costInvQtess} and \ref{fig:costirQtess}. The $\mathcal{C}^y$ is monotonically decreasing along $Q_{\text{TESS}}$, larger sizes less $C_{\textrm{grid}}$, since larger storage leads to larger flexibility and ability to capture price variations. However the investment linearly increases with size, following Eq. \ref{eq:yCosts}b, causing a tradeoff between $V_f^y$ and $\mathcal{C}^y$. The optimal size $Q^*_{\textrm{TESS}}$ is a function of the interest rate $ir$ and the unit cost $c_{\textrm{TESS}}$. The optimal capacity is shown with a red line, marking the trade-off points between $V_f^y$ and $\mathcal{C}^y$.

\begin{figure*}[tb]
    \centering
    \begin{subfigure}{\textwidth}
        \includegraphics[width=1.\linewidth]{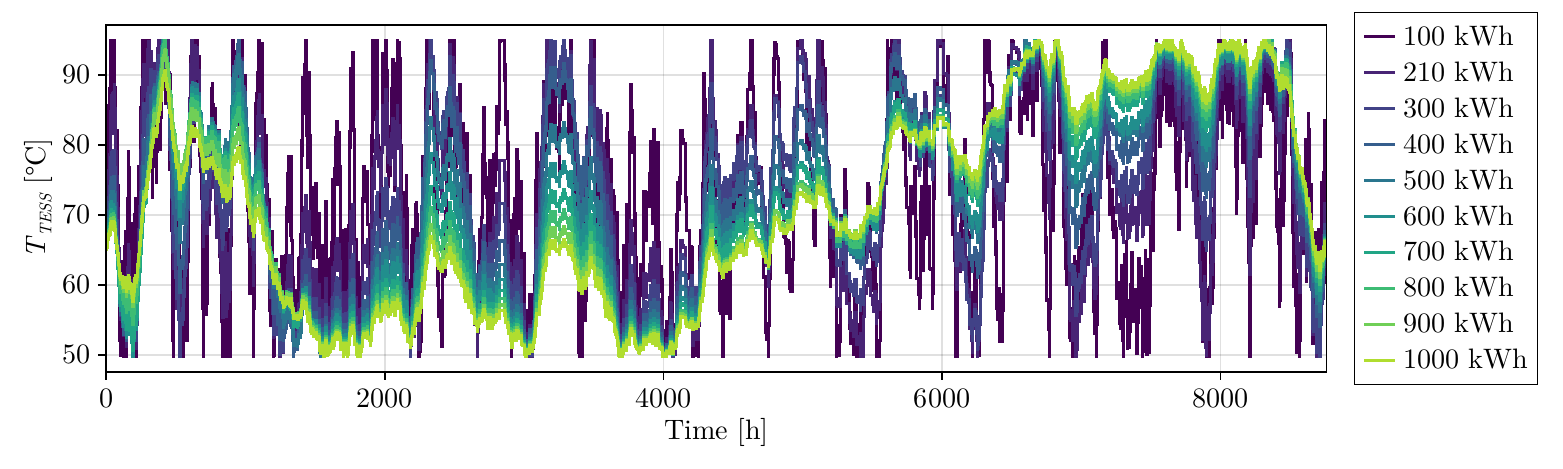} 
        \caption{Dynamic temperature response $T_{\textrm{TESS},t}$ for several $Q_{\textrm{TESS}}$ sizes.}
        \label{fig:Ttess_sizing}
        \end{subfigure}
        \medskip
        \begin{subfigure}{\textwidth}
        \includegraphics[width=1.\linewidth]{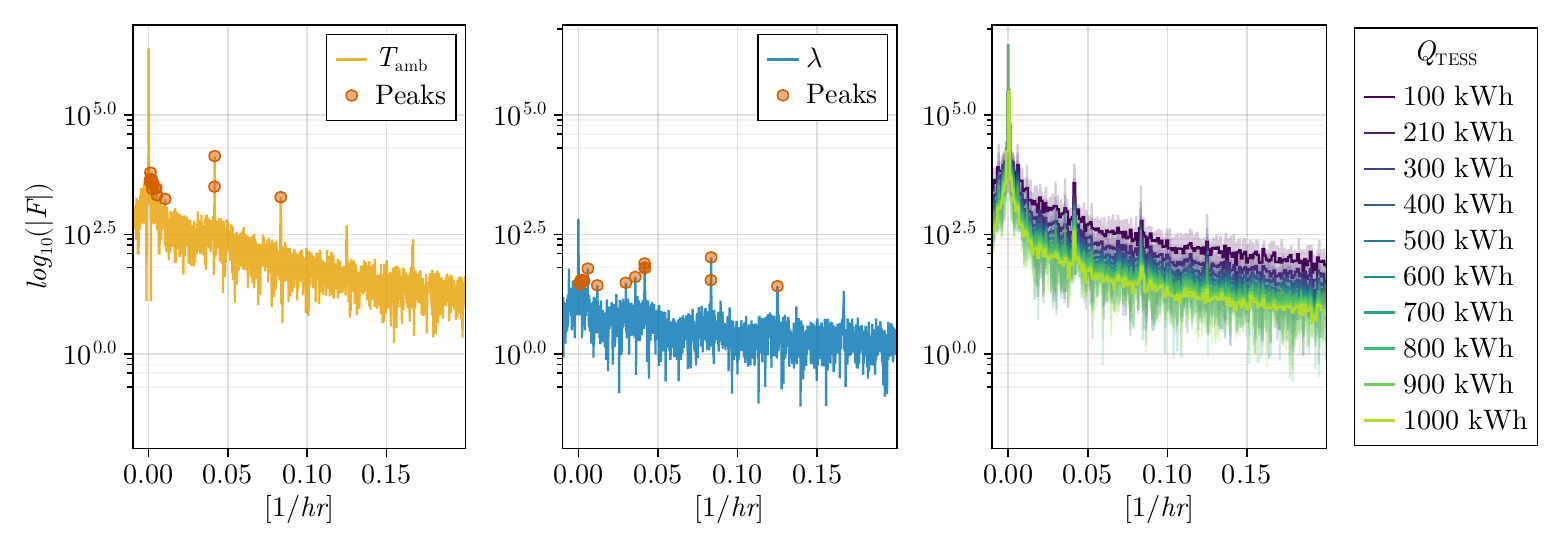}
        \caption{Discrete frequency spectrum $T_{\textrm{amb},k},\ \lambda_k\ \text{and} \ T_{\textrm{TESS},k}$ for several $Q_{\text{TESS}}$ sizes.}
        \label{fig:spectrum}
    \end{subfigure}
    \caption{Frequency analysis of the thermal storage. The ambient temperature, energy prices and \ac{tess} temperature are transformed to the frequency domain and their frequency peaks are marked.}
    \label{fig:freqSpecQtess}
\end{figure*}

Figure \ref{fig:Ttess_sizing} and \ref{fig:spectrum} present the time response and frequency response for several $Q_{\text{TESS}}$ values. Fig. \ref{fig:Ttess_sizing} shows the natural damping phenomenon of the thermal capacity with the \ac{tess} temperature, since it acts as a low-pass filter of the first order. As expected, the \ac{tess} tends to quickly charge in the first days to discharge during the winter, meaning that it is best to start the year fully charged. As spring and summer come the \ac{tess} charges capturing \ac{spv} generation and price variations to arrive to the fall fully charged. For a purely seasonal sinusoidal response of $T_{\text{TESS},t}$ the size would need to be several orders of magnitude larger, as in \cite{Darivianakis2017AManagement}. For each capacity $Q_{\textrm{TESS}}$ the output states and power dispatch are transformed to the discrete frequency domain through DFFT $(S_{a,i,t}, p_{a,i,t}) \rightarrow (S_{a,i,k}, p_{a,i,k}) $ where $k\in \mathcal{D}_k$ is the discrete frequency domain. Figure \ref{fig:spectrum} presents the frequency spectrum of the \ac{tess} state $T_{\text{TESS},k}$, and its main disturbances $T_{\text{amb},k}$ and $\lambda_k$. As expected, the disturbances $T_{\textrm{amb},k}$ and $\lambda_{k}$ present peaks at common frequencies such as 8hs, 12hs, 24hs, 36hs and so on, following the solar patterns and common routines of general users. To these exogenous disturbances the \ac{tess} responds with $T_{\textrm{TESS},k}$, which of course shows that the higher frequencies are filtered by larger \ac{tess} sizes.

Combining these two criteria an optimal \ac{tess} $Q_{\text{TESS}}^*$ and a planning horizon $H^p$ can be chosen. Assuming an interest rate $ir=6\%$ and a unit cost $c_{\textrm{TESS}}= 200$\texteuro/kWh the optimal capacity $Q^*_{\textrm{TESS}}\approx 700$ kWh. Its corresponding  slowest natural time constant is $\tau_k=60.83$days, which will be used as a planning horizon $H^p\approx 180$ days. At this point, it is important to note that the response spectrum of the \ac{tess} remains multi-modal \cite{Feng2024HybridApproach}. This differs from the common cases in the literature where purely seasonal hydrogen storage is used in the numerical examples \cite{Darivianakis2017AManagement}.

\subsection{Learning terminal set \& value function}\label{ssec:learning}

To learn the dynamic terminal set $\mathcal{S}^p_t$ the planning model in Eq. \ref{eq:planocp} is solved for scenarios $i\in \mathcal{N}_s$ as many rolling horizons $H^p$ as necessary. The terminal set is then the 95\% confidence interval of the resulting $T_{\textrm{TESS},i,t}$. To learn the value function $V_t^p(\cdot)$, each seasonal optimization of horizon $H^p$ is repeated with different initial conditions $T_{\text{TESS},t_0}$, with 10 different scenarios for each.

\begin{figure}[bth]
    \centering
    \includegraphics[width=\linewidth]{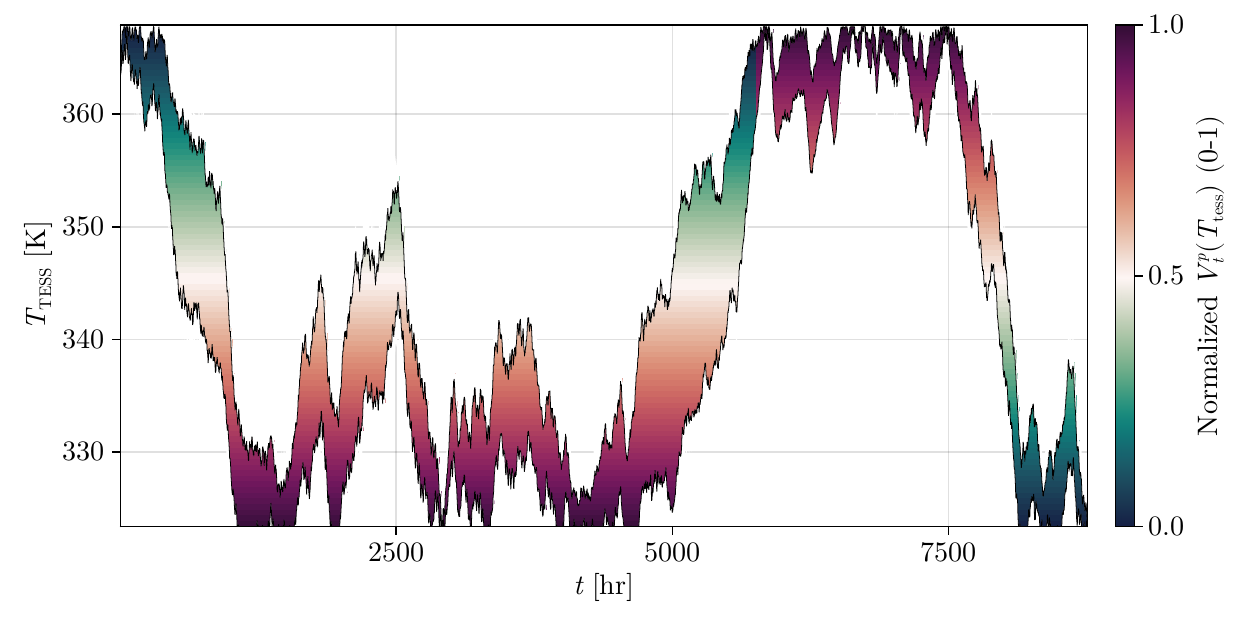}
    \caption{Dynamic terminal sets $\mathcal{S}_{t}^p$ and Value function $V_t^p(\cdot)$ for a $Q_{\text{TESS}}=700$kWh. The set $\mathcal{S}_t^p$ indicates the operation envelope where $T_{\textrm{TESS},t}$ has to finish each optimization and $V_t^p(\cdot)$ shows where in the envelope is more valuable to be. Since $V_t^p(\cdot)$ is being minimized, it is best to remain close to the green areas as possible, avoiding the red areas.}
    \label{fig:termSetVfn}
\end{figure}

The value function $V_t^p(\cdot)$ and terminal set $\mathcal{S}_t^p$ are presented in Fig. \ref{fig:termSetVfn} for three consecutive $H^p$ and the previous example of $Q^*_{\text{TESS}}=700$kWh. To create the most value the \ac{tess} temperature has to be in the vicinity of $\mathcal{S}_t^p$ and as close to the green zones as possible. On a high level, the terminal set $\mathcal{S}_t^p$ bounds the trajectories of $T_{\textrm{TESS},t}$ as time passes, guiding the short-sighted daily operational layer. The width of the terminal set $\mathcal{S}_t^p$ depends on the volatility of the random field $\mathcal{D}_W$ for that period. Naturally, periods with unpredictable weather such as the end of winter in the Netherlands or political instability impact on energy prices, produce large deviations between state trajectories $S_{a,i,t}$. Within those bounds that delimit the operation, the value function $V_t^p(\cdot)$ indicates the short-term layer where within those bounds is more valuable to be. In Fig. \ref{fig:termSetVfn} this is represented by the colormap, with the short-term \ac{mpc} having to avoid the red areas. This concept is similar to implementing a tracking \ac{mpc} where a high-level planning layer decides for an optimal reference for the \ac{tess} $T_{\textrm{TESS},t}^{\textrm{ref}}$ where the distance $\left|\left|T_{\textrm{TESS},t}-T_{\textrm{TESS},t}^{\textrm{ref}}\right|\right|_2^2$ is minimized. The advantage here lies in the flexibility of the terminal set $\mathcal{S}_t^p$ with its wider bounds, leaving room for the daily policy $X^d$ to handle short-term uncertainty.

For each horizon $H^p$ the learned $V_t^p(\cdot)$ is presented in Fig. \ref{fig:termSetVfn}. It can be seen that in the first two horizons the slope of $V_t^p(\cdot)$ is negative, meaning that within $\mathcal{S}_t^p$ is it better to discharge the \ac{tess} so that  $T_{\textrm{TESS},t}\rightarrow \underline{T}_{\textrm{TESS}}$. However, as spring and summer pass by charging the \ac{tess} and maintaining a high temperature state $T_{\textrm{TESS},t}\rightarrow \overline{T}_{\textrm{TESS}}$ has the most value for the system. For the previous example of $Q^*_{\textrm{TESS}}=700$kWh, more detailed plots can be found in Appendix \ref{app:plan}.

\subsection{Short-term Operation}

\begin{table*}[bt]
    \centering
    \begin{adjustbox}{width=1.2\textwidth}
    \begin{tabular}{cccccc}
    \hline
              & $\mathcal{C}$& $C_{\text{grid}}$ & $Q_{\text{loss}}$ & $p_T$ & $\Delta_{\text{comp}}$ \\
              & $ [\times 10^{21}]$ & [\texteuro] &[Ah] & $[\times 10^{21}]$ & [s/optimization] \\
              \hline
 \textbf{LP}& & & & & \\
    -& 2.82 $\pm$ 3.75 & 83 $\pm$ 126 & 4.45 $\pm$ 0.90 & 2.82 $\pm$ 3.75 & 6.9 $\pm$ 22.6 \\
    LP+$\mathcal{S}_{t}^p$& 1.48 $\pm$ 1.06 & 98 $\pm$ 119 & 4.83 $\pm$ 0.89 & 1.48 $\pm$ 1.06 & 7.3 $\pm$ 24.3 \\
    LP+$V_{t}^p(\cdot)$& 0.11 $\pm$ 0.29 & 66 $\pm$ 121 & 4.99 $\pm$ 0.94 & 0.11 $\pm$ 0.29 & 6.8 $\pm$ 22.9 \\
    LP+$\mathcal{S}_{t}^p\ \& \ V_{t}^p(\cdot)$& 0.01 $\pm$ 0.02 & 72 $\pm$ 123 & 5.02 $\pm$ 0.95 & 0.01 $\pm$ 0.02 & 7.0 $\pm$ 22.1 \\
    \textbf{NL}& & & & & \\
    - & 4.04 $\pm$ 7.00 & -192 $\pm$ 107 & 2.95 $\pm$ 1.46 & 4.04 $\pm$ 7.00 & 5.9 $\pm$ 31.5 \\
    \boxit{5.7in} NL+$\mathcal{S}_{t}^p$& 3.91 $\pm$ 4.87 & \textbf{-211 $\pm$ 80} & \textbf{3.29 $\pm$ 1.09}& 3.91 $\pm$ 4.87 & 7.0 $\pm$ 32.1 \\
    NL+$V_{t}^p(\cdot)$& 2.30 $\pm$ 5.30 & -207 $\pm$ 89 & 3.33 $\pm$ 1.11 & 2.30 $\pm$ 5.30 & 6.9 $\pm$ 33.2 \\
    \boxit{6in} NL+$\mathcal{S}_{t}^p\ \& \ V_{t}^p(\cdot)$& \textbf{1.80 $\pm$ 2.56}& \textbf{-209 $\pm$ 86}& \textbf{3.29 $\pm$ 1.10}& \textbf{1.80 $\pm$ 2.56}& 7.4 $\pm$ 32.7 \\
    \textbf{yOCP}& 1.05 $\pm$ 0.3 $\times 10^8$ & -6345 $\pm$ 1276 & 7.9 $\pm$ 1.6 & 1.05 $\pm$ 0.3 $\times 10^8$ & \\ 
    \hline
    \end{tabular}
    \end{adjustbox}
    \caption{Objective function components summary $\mu \pm \sigma$.}
    \label{tab:cases}
\end{table*}
To analyze the effectiveness of the proposed seasonal highway 8 \ac{empc} variations were built and simulated under $N_s=11$ scenarios, similar to the methodology used in \cite{Darivianakis2017AManagement}. The model predictive controllers are as follows:
\begin{itemize}
    \item LP, operational layer \ac{empc} from Eq. \ref{eq:sagempc}, with simplified linear dynamics $\textbf{A, B, C}$, without terminal value function $V^p_t(\cdot)$ and without terminal set $\mathcal{S}_t^p$.
    \begin{enumerate}
        \item LP + $\mathcal{S}_{t}^p$ (\textit{LP\_TS}): The LP \ac{empc} with terminal set $\mathcal{S}_t^p$ and without value function $V^p_t(\cdot)$.
        \item LP + $V_{t}^p(\cdot)$ (\textit{LP\_Vfn}): The LP \ac{empc} with value function $V^p_t(\cdot)$ and without terminal set $\mathcal{S}_t^p$.
        \item LP + $V_{t}^p(\cdot)$ \& $\mathcal{S}_{t}^p$ (\textit{LP\_Vfn\_TS}): The \ac{empc} from Eq. \ref{eq:sagempc} with simplified linear dynamics.
    \end{enumerate}
    \item NL, operational layer with Ageing-aware \ac{empc} (\textit{AGeMPC}) from Eq. \ref{eq:sagempc} with nonlinear dynamics $S^M_t$, without terminal value function $V^p_t(\cdot)$ and without terminal set $\mathcal{S}_t^p$ \cite{Slaifstein2026SequentialMPC}. 
    \begin{enumerate}
        \item NL + $\mathcal{S}_{t}^p$ (\textit{NL\_TS}): The NL \ac{empc} with terminal set $\mathcal{S}_t^p$ and without value function $V^p_t(\cdot)$.
        \item NL + $V_{t}^p(\cdot)$ (\textit{NL\_Vfn}): The NL \ac{empc} with value function $V^p_t(\cdot)$ and without terminal set $\mathcal{S}_t^p$.
        \item NL + $V_{t}^p(\cdot)$ \& $\mathcal{S}_{t}^p$: Seasonally \& Ageing-aware \ac{empc} (\textit{SAGeMPC}) from Eq. \ref{eq:sagempc}.
    \end{enumerate}
    \item Yearly Optimal Control (\textit{yOCP}) with simplified linear dynamics $\textbf{A, B, C}$.
\end{itemize}


The main performance metrics of the controllers are presented in Table \ref{tab:cases}. The parameters are the total operational cost $\mathcal{C}$, the grid cost $C_{\textrm{grid}}$, the battery capacity lost $Q_{\text{loss}}$, the thermal comfort penalty $p_T$ and the computational time $\Delta_{\textrm{comp}}$. Regarding, computational time $\Delta_{\text{comp}}$, all controllers have similar means (from 5.9 to 7.4). Whereas deviations seems to be separated depending on the dynamics, with \textit{LP} controllers having 22-24s and the \textit{NL} controllers having 31-33s. This is expected behavior for  Of course, the soft constraint of the inside temperature $p_T$ takes up most of the operational cost $\mathcal{C}$ due to the high value of penalty weight $w_T$. This is caused by the fact that within a single optimization step a small deviation has to be largely penalized with $w_{T}$ to ensure thermal comfort. The proposed \ac{empc}s reach values between $\times 10^{19}$ to $\times 10^{21}$. However, this does not mean the inhabitants are uncomfortable since the penalty $p_T$ measures both the magnitude of the excursion and its duration. In particular, the duration of the excursions reaches its highest value of approximately 10\% of the total year for the \textit{LP\_Vfn\_TS} controller, with the rest of the benchmarks being below that. Thermal discomfort would be a value closer to the $\times 10^{30}$, with highly frequent and large excursions outside of $\underline{T}_{\text{in},t}, \overline{T}_{\text{in},t}$. Summarizing, the \textit{LP} controllers achieve the best thermal comfort among the proposed \ac{mpc}s, with the best \textit{NL} controller being the \textit{SAGeMPC}.

However, achieving the lowest $C_{\text{grid}}$ is the primary goal the best controllers are the nonlinear controllers due to their ability to fully exploit the \ac{bess} and \ac{hp}. In this regard, the \textit{NL\_TS} with the terminal set $\mathcal{S}_t^p$ achieves the best mean grid cost, the second best degradation and the second worst comfort penalty $p_T$. Among the nonlinear \ac{empc}, the full \textit{SAGeMPC} controller presents the most balanced performance with the second best grid cost $C_{\textrm{grid}}$, better capacity fade $Q_{\textrm{loss}}$ than all linear versions, the best nonlinear thermal comfort $p_T$ and a mean computational time on par with the linear \ac{empc}s. If the thermal comfort is not a priority then \textit{NL\_TS} is the best alternative.

\begin{figure*}[!bt]
    \centering
    \includegraphics[width=\linewidth]{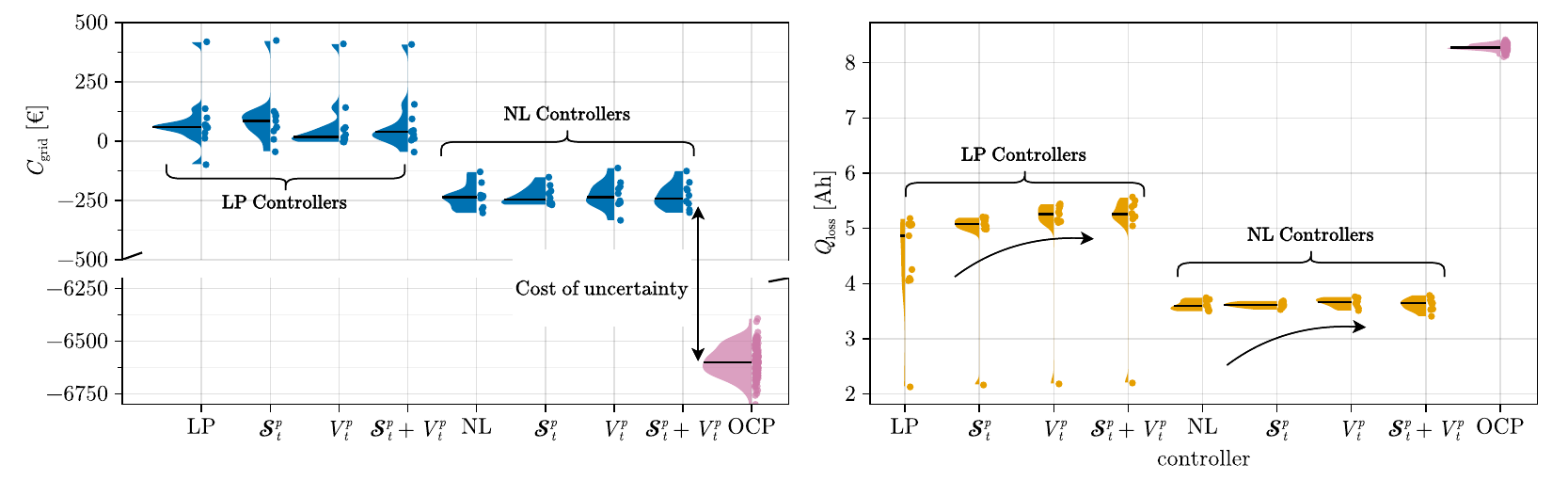}
    \caption{Distributions of grid cost $C_{\text{grid}}$ and capacity fade $Q_{\text{loss}}$ for each controller. From linear \textit{LP} to nonlinear \textit{NL} and from plain to fully seasonally aware. The reader must not overlook the broken y-axis in the $C_{\text{grid}}$.}
    \label{fig:Qloss}
\end{figure*}

Besides these mean metrics, it is also interesting to stop and analyze the full distributions of $C_{\text{grid}}$ and $Q_{\text{loss}}$ and check its relationship with the volatility of inputs $\mathcal{D}_W$. Both distributions are presented in Fig. \ref{fig:Qloss}. Starting with the grid cost, it is clear that the nonlinear controllers consistently beat the linear controllers, at the expense of thermal comfort $p_T$. Reversely, the \textit{LP} controllers maintain a better thermal comfort $p_T$ at the expense of spending more on grid cost $C_{\text{grid}}$. In particular, the use of the terminal set $\mathcal{S}_{H^d}^p$ shapes the distributions of the controllers in which is used (\textit{NL\_TS} and \textit{NL\_Vfn\_TS}). Lastly, the linear \textit{yOCP} represents the long-term optima baseline, achieving the best across all of its optimized metrics (which do not include $Q_{\textrm{loss}}$). Naturally, the presented gap of almost \texteuro6000 in favor of the \textit{yOCP} is the cost to pay for using the daily horizon $H^d$ and adjusting against short-term uncertainty.

On the battery degradation side, the two most clear takeaways are: the nonlinear \textit{AGeMPC} has the best degradation control and the excellent $C_{\textrm{grid}}$ distribution of the \textit{yOCP} comes at the expense of almost doubling battery degradation with respect to the \textit{AGeMPC}. For all controllers seasonally steering the \ac{tess} comes at the expense of higher battery degradation outcomes. In other words, a trade-off between degradation control and seasonal coordination appears when using terminal sets and value functions. The \textit{LP} controllers appear to be more sensitive to this phenomenon, whereas battery degradation control is kept tightly in the \textit{NL} controllers. Meaning that seasonal \ac{tess} control does not affect ageing control when accounted in the decision making optimizer. This is related to the fact that the data-driven seasonal components are learned from policies mainly minimizing grid cost, unaware of battery ageing.

\begin{figure}[bth]
    \centering
    \includegraphics[width=0.8\linewidth]{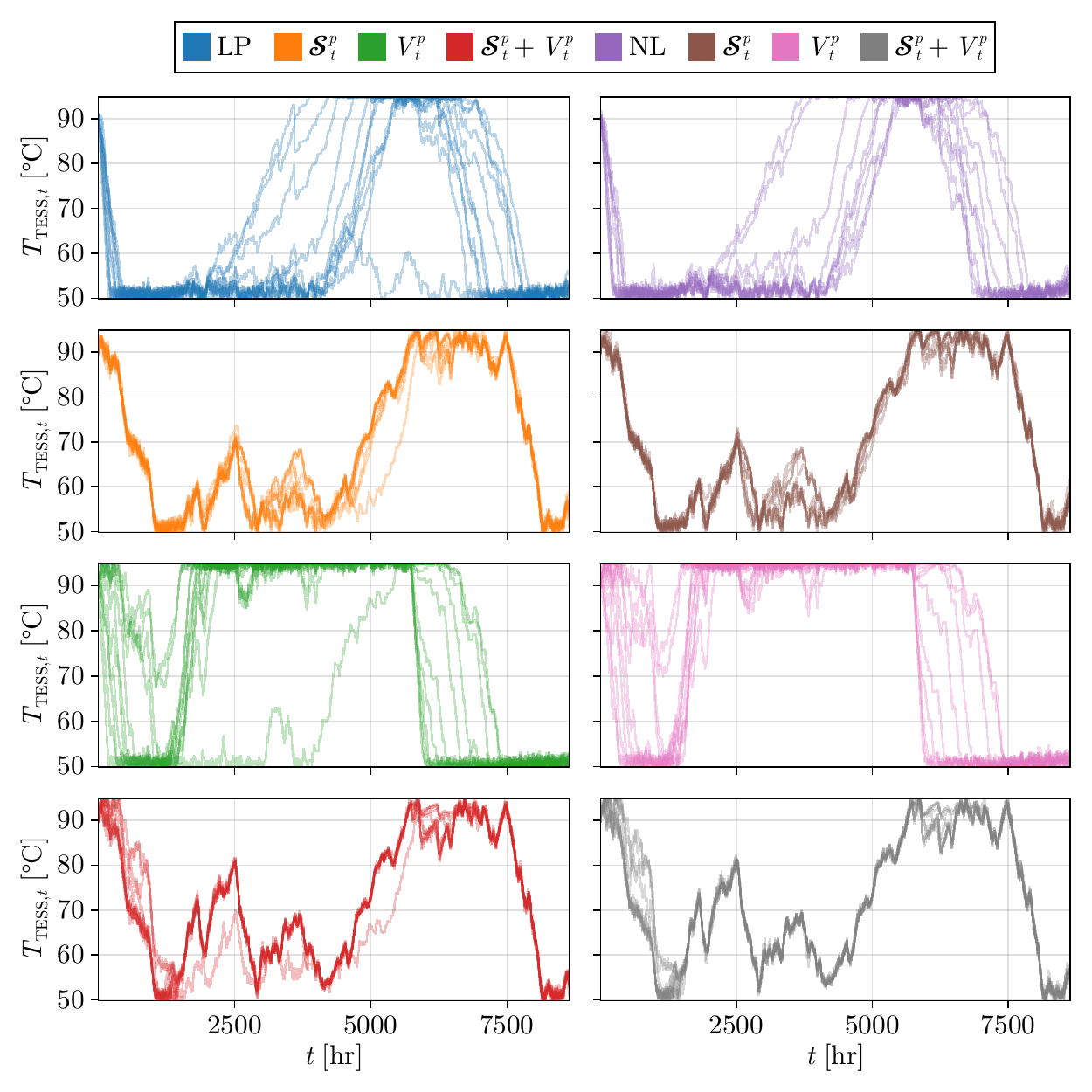}
    \caption{\ac{tess} state trajectories $T_{\text{TESS},t}$ for each of the different controllers under different scenarios. On the left the \textit{LP} controllers, on the right the \textit{NL} controllers. From top to bottom, the plain controllers, with the added dynamic terminal set $\mathcal{S}_{t}^{p}$, with the value function $V_{t}^p$,  with full seasonal components.}
    \label{fig:tessYear}
\end{figure}

It is also interesting to analyze the state trajectories of the \ac{tess}. The \ac{tess} temperature across scenarios and controllers $T_{\textrm{TESS},i,t}$ is presented in Fig. \ref{fig:tessYear}. Besides a few small differences between linear and nonlinear controllers, the overall behavior of the thermal storage is mainly affected by the combination of $\mathcal{S}_t^p$ and $V_t^p(\cdot)$. In the myopic \ac{empc}s, \textit{LP} and \textit{NL}, the \ac{tess} charges quickly during the spring-summer, depending on the disturbances $W_{i,t+1}$. In this controller, the \ac{tess} is charged when there is a price arbitrage opportunity in the horizon (large $\Delta \lambda_t$) and the building and \ac{bess} are close to their upper bound ($T_{\textrm{in},t}\rightarrow \overline{T}_{\textrm{in},t}$ and $SoC_{\textrm{BESS},t}\rightarrow \overline{SoC}_{\textrm{BESS},t}$). Otherwise, the controller chooses to deplete the \ac{tess}. This is prevented in the controllers with terminal set $\mathcal{S}_t^p$, \textit{LP\_TS} and \textit{NL\_TS}, since it has to remain close to the set $\mathcal{S}_t^p$ defined by the thermal planning optimal. On the contrary the value function $V_{t}^p$ incentivizes greedy behavior (minimizing short-term grid cost). However, just as in \cite{Darivianakis2017AManagement}, its addition on itself does not effectively steer the  system towards seasonal optimality when it is not combined with the \ac{tess} terminal set $\mathcal{S}_t^p$. This limitation is addressed by Brosius \textit{et al} \cite{Brosius2026MitigatingFunctions}, by continuously updating the value function at each timestep $t$. Balance is achieved once the two components are used at the same time, effectively passing valuable seasonal information to the short-term \ac{empc} operation.

Recapping, the data-driven seasonal components $\mathcal{S}_t^p$ and $V_t^p(\cdot)$ can effectively steer nonlinear \ac{empc} towards seasonal optimality. Among all the presented controllers, the most balanced option is the \textit{SAGeMPC} which combines detailed nonlinear dynamic models with the presented seasonal components. The \textit{SAGeMPC}:
\begin{itemize}
    \item has the second lowest mean grid cost $C_{\text{grid}}$ within \ac{mpc}s, $\mathbb{E}[C_{\text{grid}}]=-\text{\texteuro}209$.
    \item has the second best mean battery degradation $Q_{\text{loss}}$, $\mathbb{E}[Q_{\text{loss}}]=3.29$Ah.
    \item presents an acceptable thermal comfort and computational time of $\mathbb{E}[\Delta_{\text{comp}}]=7.4$s.
\end{itemize}

\section{Discussion}

Beyond the results, the present framework has to be understood in the broader context of the field. As discussed in Section \ref{ssec:learning}, this work improves on the standard hierarchical architecture of offline planning simulations and online tracking \ac{mpc} because it is too restrictive under high volatility $\mathcal{D}_W$. Moreover, for some applications there is not enough data to come up with long term scenarios, especially in very specific commercial or industrial applications where field measurements are critical. Thus, it is extremely necessary to stay flexible during daily operation, favoring economic \ac{mpc} instead of tracking \ac{mpc}. With this in mind choosing suitable terminal components (value functions and terminal sets) becomes critical when approaching \ac{ems} design in a holistic manner.

Regarding the effectiveness of the terminal value function $V_t^p(\cdot)$: in the presented framework its purpose is only to guide the \ac{tess} within the terminal set $\mathcal{S}_t^p$, it does not provide any useful information outside of it. This is because it was learned on the planning scenarios where $H^p > H^d$, where naturally $T_{\textrm{TESS},t}\in\mathcal{S}_t^p$. Naturally, since the value function $V_t^p(\cdot)$ is only a linear regressor other regressor classes should be explored, in particular deep learning/MILP classes. An alternative is presented by Brosius \cite{Brosius2026MitigatingFunctions} where no terminal set is used and the value functions are updated on-line during operation. In synthesis, to steer short-sighted \ac{empc} the seasonal information approach has to be consistent. Either use the duo of $\mathcal{S}_t^p$ and $V_t^p(\cdot)$ or cost-to-go functions updated online or multi-horizon approaches \cite{Darivianakis2017AManagement, Brosius2026MitigatingFunctions, Behrunani2024DistributedHubs}. However, what is interesting about the presented approach is its explainability and usefulness to interpret different time-scale energy storage and its flexibility in terms of the dynamic models used (linear or nonlinear).

While the dynamic terminal sets $\mathcal{S}_t^p$ and value functions $V_t^p(\cdot)$ are pre-computed offline using multi-year historical stochastic scenarios, severe climate anomalies or unannounced structural energy price events could shift the boundary of viable operational trajectories. To preserve robustness without compromising real-time computational efficiency during daily economic dispatch ($H^d = 24\text{ h}$), future implementations can introduce a background sliding-window re-synthesis. By re-evaluating the dynamic terminal bounds over an updated $H^p$ window at low frequency (e.g., monthly), the framework dynamically adapts long-term steering boundaries and value functions to multi-year exogenous shifts while maintaining short-horizon online control.

The integration of continuous electrochemical aging models within \textit{SAGeMPC} achieves a $26\text{-}34\%$ reduction in BESS capacity loss compared to linear \ac{mpc} formulations. However, enforcing long-term thermal steering ($\mathcal{S}_t^p\ \& \ V_t^p(\cdot)$) leads to a modest increase in capacity loss ($3.29\text{Ah}$) relative to unconstrained non-linear \ac{mpc} ($2.95\text{Ah}$) over the evaluation period $T$. This $0.34\text{Ah}$ penalty occurs because maintaining TESS state trajectories within winter energy-reserve corridors requires additional heat pump operation during transition seasons, driving higher peak discharge rates in the BESS. This small operational degradation cost represents a necessary trade-off to eliminate seasonal depletion risks and guarantee long-term thermal self-sufficiency.


Finally, regarding the cost of uncertainty, i.e. the cost difference between \textit{yOCP} and \ac{empc}, the large deviation comes from the difference between $H^d$ and $H^y$. The reader might notice that when analyzing against \cite{Darivianakis2017AManagement} the cost difference is not that large. However, the long-term storage used in \cite{Darivianakis2017AManagement} is hydrogen and of a larger ratio against the \ac{bess} than in the system presented in Section \ref{sec:results}  (around 8 thousand times larger). Thus, it is intuitive to assume that the difference between horizons ($H^y>>H^p>>H^d$) and the storage sizes ratio drives the cost difference between \textit{yOCP} and \ac{empc} alternatives. In other words, as the ratio between long-term and short-term storage grows, the planning horizon converges to the yearly horizon ($H^p\rightarrow H^y$) and the difference between seasonally-aware \ac{empc} and \ac{ocp} shrinks ($\mathcal{C}^{\text{MPC}}\rightarrow\mathcal{C}^{\text{yOCP}}$). This is left out of the scope of this work since the financial parameters used ($ir$ and $c_{\textrm{TESS}}$) are already optimistic, making even larger \ac{tess} sizes unrealistic. Theoretical work on this line is highly encouraged for future research.

\section{Conclusions \& Future works}
\label{sec:conclusions}

This work presents a data-driven framework to size, plan, and operate residential energy hubs that combine long-term thermal storage with short-term electrical storage. Frequency response is combined with financial information to jointly determine the optimal \ac{tess} capacity $Q^*_{\textrm{TESS}}$ and the minimum planning horizon $H^p$ needed to capture its seasonal behavior. This avoids the need for intractable yearly stochastic optimizations at the planning stage. For the presented case study, this resulted in an optimal capacity $Q^*_{\textrm{TESS}}\approx700$kWh and a planning horizon $H^p\approx180$ days, driven by the slowest natural time-constant of the \ac{tess} response to ambient temperature $T_{\text{amb},t}$ and price disturbances $\lambda_t$. The optimal capacity is used in thermal planning simulations to learn seasonal patterns.

The learned dynamic terminal set $\mathcal{S}_t^p$ and terminal value function $V_t^p(\cdot)$ were shown to effectively link this seasonal planning layer to a short-sighted daily \ac{empc}. Neither component is sufficient on its own: the value function alone tends to induce greedy, short-term behavior, while the terminal set alone constrains the \ac{tess} trajectory without providing economic guidance; only their combination reliably steers the daily policy towards seasonal optimality. Among all tested controllers, the nonlinear, ageing- and seasonally-aware \textit{SAGeMPC} offered the most balanced performance, achieving the second best mean grid cost of any \ac{empc} ($\mathbb{E}[C_{\text{grid}}]=-$\texteuro209), a mean battery degradation lower than all linear alternatives ($\mathbb{E}[Q_{\text{loss}}]=3.29$Ah), an acceptable thermal comfort and computational time of $\Delta =7.4s$ aligned even with linear \ac{empc} benchmarks. Compared to the long-horizon \textit{yOCP} benchmark, \textit{SAGeMPC} tries to close the cost gap between short-term optima and long term optimality at a fraction of the computational cost, with the remaining $\sim$\texteuro6,000 difference attributable to the unavoidable mismatch between the daily operational horizon $H^d$ and the yearly horizon $H^y$.

Future work will extend the framework to include \ac{ev} charging and bidirectional (V2G) flexibility, investigate on-line updating of the seasonal components during operation to relax the dependency on the offline planning optimizations and further characterize how the storage-size ratio between long- and short-term assets governs the convergence of the planning horizon $H^p$ towards the yearly horizon $H^y$, particularly for storage technologies with larger seasonal-to-daily capacity ratios such as hydrogen.


\section{Acknowledgment}

The project was carried out with a Top Sector Energy subsidy from the Ministry of Economic Affairs and Climate, carried out by the Netherlands Enterprise Agency (RVO). The specific subsidy for this project concerns the MOOI subsidy round 2020.

\appendix
\section{Device Models}\label{app:models}
In this Appendix a summary of the models used in the paper are presented. For the electrical power balance:
\begin{equation}
   P_{\text{PV},t}+P_{\text{BESS},t}+ P_{\text{grid},t}=P_{\text{load},t}^{\textrm{e}}+P_{\text{HP},t}
\end{equation}

The dynamical model of the building is:
\begin{subequations}\label{eq:Tin}
    \begin{align}
        & T_{\text{in},t+1}  = T_{\textrm{in},t}+\frac{\Delta t}{C_b + V_b \rho_{\textrm{air}} C_{\textrm{air}}}  \Big( \dot{Q}_{\textrm{ir},t} + \dot{Q}_{\textrm{TESS},t}^{\textrm{D}} +\notag \\
        & \qquad \qquad \dot{Q}_{\textrm{HP},t}^{\textrm{D}} - \dot{Q}_{\textrm{loss},t} \Big)\\
        & \dot{Q}_{\textrm{ir},t}= w_b s_b G_{\textrm{ir},t} \sum_{s=2}^3 A_s\\
        & \dot{Q}_{\textrm{loss},t}= \dot{Q}_{\textrm{cond},t}+\dot{Q}_{\textrm{vent},t} \\
        & \dot{Q}_{\textrm{vent},t}= C_{\textrm{air}} \rho_{\textrm{air}} V_b r_b  (T_{\text{in},t} - T_{\text{amb},t}) \\
        & \dot{Q}_{\textrm{cond},t}= (T_{\text{in},t} - T_{\text{amb},t}) \sum_{s=1}^{\mathcal{S}} U_s A_s
    \end{align}
\end{subequations}
For the heat pump two models are used. The complete model for the \ac{empc} of Eq. \ref{eq:sagempc} with nonlinear coefficient of performance is:
\begin{subequations}\label{eq:hp}
    \begin{align}
        & P_{\text{HP},t}=\frac{\dot{Q}_{\text{HP},t}^{\textrm{TESS}}}{COP_{\text{HP},t}^{\text{TESS}}}+\frac{\dot{Q}_{\text{HP},t}^{\textrm{D}}}{COP_{t}^{\text{D}}}\,, \\
    & T_{\text{HP},t}^{\textrm{D, in}} = T_{\text{HP},t}^{\textrm{D, out}} - \frac{\dot{Q}_{\text{HP}}^{\text{D}}}{\eta_{\text{HP}} \dot{m}_f c_f} \\
    & T_{\text{HP},t}^{\textrm{TESS, in}} = T_{\text{HP},t}^{\textrm{TESS, out}} - \frac{\dot{Q}_{\text{HP}}^{\text{TESS}}}{\eta_{\text{HP}} \dot{m}_f  c_f}\\
    & COP_t^{\text{D}} = 7.90471 e^{-0.024 (T_{\textrm{HP},t}^{\text{D, in}} - T_{\text{amb},t})} \\
    & COP_t^{\text{TESS}} = 7.90471 e^{-0.024 (T_{\textrm{HP},t}^{\text{TESS, in}} - T_{\text{amb},t})} 
    \end{align}
\end{subequations}
\noindent{in the planning models, Eqs. \ref{eq:yearOpt} and \ref{eq:planocp}, the coefficient is fixed to $COP^{\text{D}}_t=COP^{\text{TESS}}_t=3$.}
Finally, for the thermal energy storage its dynamics are modelled linearly as:
\begin{subequations}\label{eq:TESS}
    \begin{align}
       & T_{\text{TESS}, t+1} =  T_{\text{TESS}, t} + \frac{\Delta  t}{m_{\text{TESS}}  c_{\text{TESS}}} \left(  \dot{Q}_{\text{HP},t}^{\textrm{TESS}} - \eta \dot{Q}_{\text{TESS},t}^{\text{D}} \right) \\ 
        & SoC_{\text{TESS},t}=\frac{T_{\textrm{TESS},t}-\underline{T}_{\textrm{TESS}}}{\overline{T}_{\textrm{TESS}}-\underline{T}_{\textrm{TESS}}}
    \end{align}
\end{subequations}
    

The complete model for battery performance and degradation is:
\begin{subequations}\label{eq:batt}
    \begin{align}
        & SoC_{b,t+1}=SoC_{b,t}-\frac{\Delta t}{Q_{b,t}3600}\eta_c i_{b,t} \\
        & P_{b,t}= N_{s,b}N_{p,b} v_{t,b,t}  i_{b,t} \\
        & i_{R_1,b,t+1}=e^{-\frac{\Delta t}{R_{1,b}C_{1,b}}}i_{R_1,b,t}+ \left( 1-e^{-\frac{\Delta t}{R_{1,b}C_{1,b}}} \right) i_{b,t} \\
        & OCV_{b,t} = OCV_{p,b,t}(SoC_{b,t}) - OCV_{n,b,t}(SoC_{b,t})\\
        & v_{t,b,t}=OCV_{b,t}-i_{R_1,b,t}R_{1,b}-i_{b,t}R_{0,b} \\
        & i_{\textrm{SEI},b,t} = \frac{k_{\textrm{SEI},b}e^{\frac{-E_{\textrm{SEI},b}}{RT}}}{n_{\textrm{SEI}}(1+\lambda_{b}\beta_{b})\sqrt{t_{0,b}+t}} \\
        & \beta_{b} = e^{\frac{n_{\textrm{SEI}}F}{R T} \left(\eta_{k,b}+OCV_{n,b,t}-OCV_s \right)} \\
        & z_{b,t} = SoC_{b,t} (z_{100\%}-z_{0\%})+z_{0\%} \\
        & \eta_{k,b,t}=\frac{2 R T}{F} \text{sinh}^{-1}\left(\frac{i_{b,t}}{n_{\textrm{SEI}} a_s A L_n i_0}\right) \\
        & i_{\textrm{AM},b,t} = k_{\textrm{AM},b} e^{\frac{-E_{\textrm{AM},b}}{R T}} SoC_{b,t} |i_{b,t}| Q_{b,0}\\
        & i_{\text{loss},\ b,t} = i_{\textrm{SEI},b,t} + i_{\textrm{AM},b,t} \\
        & Q_{b,t+1}=Q_{b,t} -\frac{\Delta t}{3600} i_{\text{loss},b,t}
    \end{align}
\end{subequations}

\section{Thermal Planning simulations}\label{app:plan}

\begin{figure}[tb]
    \centering
    \includegraphics[width=\textwidth]{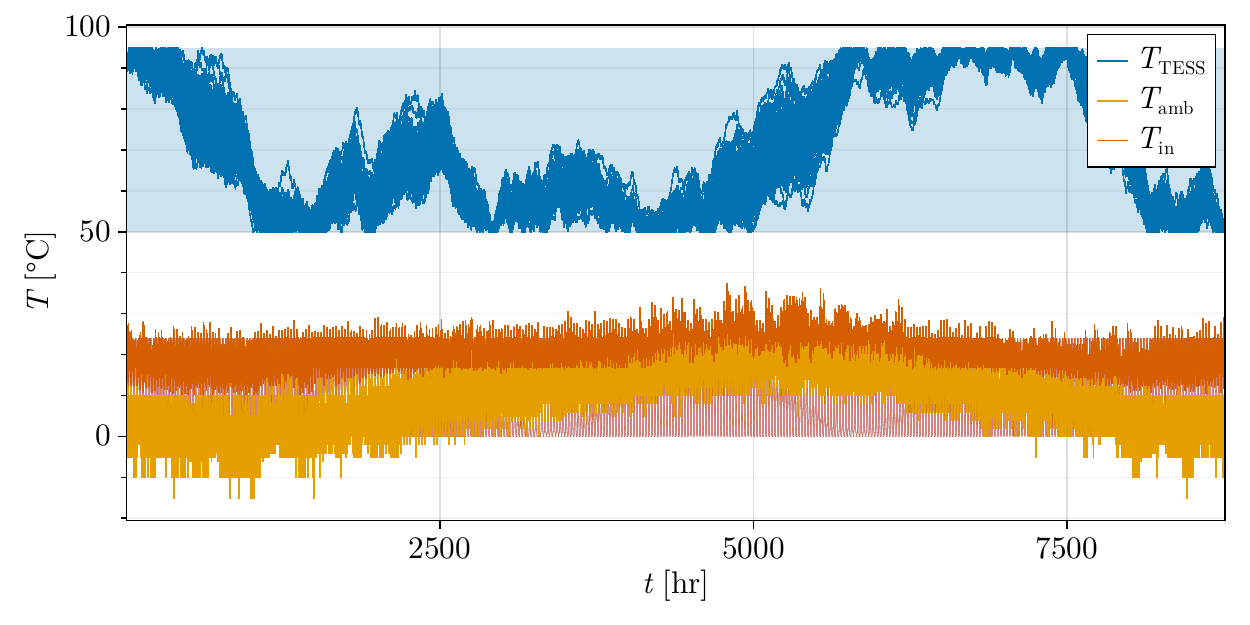}
    \caption{Thermal planning simulation trajectories of $T_{\text{TESS},i,t}$ and $T_{\textrm{in},i,t}$ for $Q^*_{\text{TESS}}=700$kWh, $H^p=180$days and $N_s=100$.}
    \label{fig:planningSoC}
\end{figure}

\begin{figure}[bth]
    \centering
    \includegraphics[width=\textwidth]{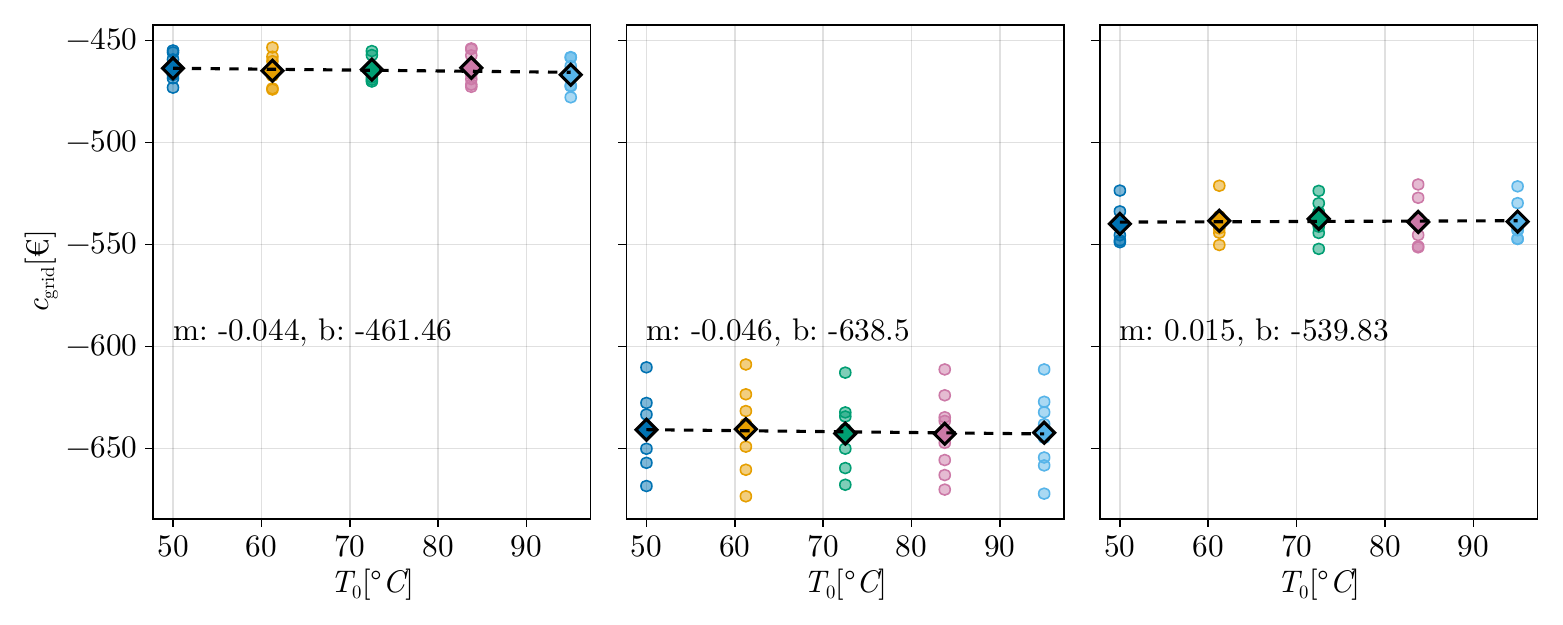}
    \caption{Learned linear Value functions $V_t^p(\cdot)$ for each planning horizon $H^p$, using five initial conditions $T_{0}$ and $N_s=10$ scenarios for each.}
    \label{fig:learnVfn}
\end{figure}

This Appendix presents intermediate results for the thermal planning simulations. For the terminal set of the \ac{tess} $\mathcal{S}_t^p$, Fig. \ref{fig:planningSoC} presents the trajectories of the building and \ac{tess} under scenario simulation. The trajectories of $T_{\text{TESS},i,t}$ define the envelope of operation in response to the scenarios $T_{\text{amb},i,t}$. From the terminal value function simulations the resulting value function $V_t^p(\cdot)$ of each planning horizon $H^p$ are presented in Fig. \ref{fig:learnVfn}.

 \bibliographystyle{elsarticle-num} 
 \bibliography{references}
 \biboptions{sort&compress}





\end{document}